\documentclass{aa}
\usepackage{graphicx}
\usepackage{txfonts}
\usepackage{subcaption}         
\usepackage{placeins}           
\usepackage[colorlinks=true, linkcolor=blue, citecolor=blue, filecolor=blue, urlcolor=blue]{hyperref}
\usepackage{booktabs}
\graphicspath{{./plots_paper/}}

\begin{document}

\title{Forecasts of effects of beam systematics and deprojection\\ on the third-generation ground-based cosmic microwave background experiment}
\author{Jiazheng Dou\inst{1,2}\thanks{\email{doujzh@mail.ustc.edu.cn}}
\and Jiakang Han\inst{3,4}
\and Wen Zhao\inst{1,2}\thanks{\email{wzhao7@ustc.edu.cn}}
\and Bin Hu\inst{5,6}\thanks{\email{bhu@bnu.edu.cn}}
}
\institute{Department of Astronomy, University of Science and Technology of China, Chinese Academy of Sciences, Hefei, Anhui 230026, People’s Republic of China
\and School of Astronomy and Space Sciences, University of Science and Technology of China, Hefei 230026, People’s Republic of China
\and Dipartimento di Fisica, Universit\`a degli Studi di Torino, Via P.\ Giuria 1, 10125 Torino, Italy
\and INFN -- Istituto Nazionale di Fisica Nucleare, Sezione di Torino, Via P.\ Giuria 1, 10125 Torino, Italy
\and Institute for Frontier in Astronomy and Astrophysics, Beijing Normal University, Beijing, 102206, People’s Republic of China\label{inst5}
\and Department of Astronomy, Beijing Normal University, Beijing 100875, People’s Republic of China}


\abstract{The ground-based cosmic microwave background (CMB) experiments are susceptible to various instrumental errors, especially for $B$-mode measurements. The difference between the response of two polarized detectors, referred to as the beam mismatch, would induce a $T\rightarrow P$ leakage when the detector pair is differenced to cancel the unpolarized signal. We applied the deprojection technique on the time-ordered mock data to mitigate the systematic contamination caused by beam mismatches by assuming the third-generation ground-based CMB experiment (S3). Our results show that the deprojection effectively recovered the input power spectra. We adopted the Needlet ILC (NILC) and constrained ILC (cILC) methods to reconstruct the foreground-cleaned $TEB$ maps, and we evaluated the level of residual systematic errors after the foreground cleaning pipeline by comparing the power spectra between the systematics-added data after deprojection and the systematics-free data. The results show that the residual beam systematics cleaned by deprojection do not bias the CMB measurements of the $T$, $E$, and $B$ modes nor the CMB lensing reconstruction {or the estimation of the tensor-to-scalar ratio} under the S3 sensitivity.
}

\titlerunning{Forecasts of effects of beam systematics and deprojection on S3}
\authorrunning{J. Dou et al.}
\maketitle

\section{Introduction}
Precise measurements of cosmic microwave background (CMB) polarization are particularly essential for modern cosmology and for exploring new physics beyond the $\Lambda$CDM model \citep{aghanimPlanck2018Results2020a}. The detection of the primordial $B$-mode polarization would reveal the mystery of the cosmic inflation era. The amplitude of the primordial $B$ modes is commonly described by the ratio of the tensor power spectrum to the scalar power spectrum at some pivot scale, called the tensor-to-scalar ratio. Successive ground-based CMB experiments over the past decade have made great efforts to increase their sensitivities to enable a significant detection of the tensor-to-scalar ratio, such as BICEP/Keck {Array} \citep{collaborationBICEPKeckXIII2021}, {the Atacama Cosmology Telescope (ACT)} \citep{louisAtacamaCosmologyTelescope2017}, {the Cosmology Large Angular Scale Surveyor (CLASS)} \citep{Thomas_2014,Harrington_2016}, and {the South Pole Telescope (SPT)} \citep{sayreMeasurementsBmodePolarization2020}. The current best constraint on the tensor-to-scalar ratio is $r<0.032$ (95\% C.L.) and was obtained from the combination of Planck, BAO, and BICEP/Keck data \citep{tristramImprovedLimitsTensortoscalar2022}. Future CMB experiments such as {the Ali CMB Polarization Telescope (AliCPT)} \citep{Li:2017drr,Li:2018rwc}, the Simons Observatory \citep{thesimonsobservatorycollaborationSimonsObservatoryScience2019}, {the Q\&U Bolometric Interferometer for Cosmology (QUBIC)} \citep{mennellaQUBICExploringPrimordial2019}, LiteBIRD \citep{hazumiLiteBIRDSatelliteStudies2019}, and CMB-S4 \citep{abazajianCMBS4ScienceBook2016,CMB-S4:2022ght,collaborationCMBS4ForecastingConstraints2022} will continue to pursue the target of higher sensitivities on constraining $r$.

Precise measurements of the CMB polarization are challenging not only because it is weaker than the instrumental noise, but also because of complicated contaminants from astrophysics and instruments. Diffuse galactic foregrounds such as synchrotron and thermal dust emit strong polarized radiation at microwave frequency bands with a significant amplitude relative to the CMB polarization signal. Component separation methods are implemented on the multifrequency sky observations to remove the foregrounds (mostly) based on the fact that CMB and foregrounds have different spectral energy distributions (SED). Our imperfect knowledge of the actual optical response of the telescopes results in instrumental systematic effects, such as the unknown difference between the response of two polarized detectors that might transform a fraction of the CMB temperature modes into polarization modes. This is called the $T\rightarrow P$ leakage. The leakage must be mitigated especially because the temperature modes and the polarization modes peak at similar angular scales so that the leakage of the former has a higher magnitude than the signal of the latter. Additionally, the gravitational lensing effect where the matter field gravitationally deflects the CMB photons can covert part of $E$ modes into $B$ modes at intermediate and small scales, giving rise to lensing $B$ modes that dominate the primordial $B$ modes \citep{Blanchard1987,zaldarriagaGravitationalLensingEffect1998}. Fortunately, the lensing convergence can be reconstructed using either internal CMB anisotropies or external large-scale structure (LSS) tracers such as galaxy surveys, 21 cm observations, and measurements of the cosmic infrared background (CIB) \citep{Hu:2001tn,Hu:2001kj,Okamoto:2003zw,Carron:2017mqf,Maniyar:2021msb,planckcollaborationPlanck2018Results2020}. 

We consider a typical configuration for a small-aperture ground-based CMB telescope. Specifically, we assumed a total of 7,000 polarized detectors, evenly split between the $95$ GHz and $150$ GHz frequency bands. Each detector had a noise equivalent temperature (NET) of \(350 \, \mu\text{K}\sqrt{s}\). The full width at half maximum (FWHM) beam sizes were assumed to be $19$ arcminutes and $11$ arcminutes for the $95$ GHz and $150$ GHz bands, respectively. This setup reflects the typical design of a third-generation ground-based CMB experiment (S3) \citep{abazajianCMBS4ScienceBook2016}.
We focus on exploring the impact of the beam mismatch, which is a difference in the beam shape or beam center between the two detectors of a detector pair that leads to the $T\rightarrow P$ leakage. An analysis technique called the deprojection was exploited to filter this contamination out \citep{bicep2collaborationBICEP2IIIInstrumental2015,sheehyDeprojectingBeamSystematics2019}. We used the deprojection algorithm to handle the time-ordered data {(TOD)} with the beam systematics, and we evaluated the effects of undeprojected residual systematics on the follow-up steps of our whole pipeline, including component separation, lensing reconstruction, {and the estimation of the tensor-to-scalar ratio}. The residual systematic error was found to be negligible given the sensitivity of S3.

The paper is organized as follows. In Sect.~\ref{sec:bs} we first review the formalism of beam mismatches modeled by six coefficients. Then, we introduce our deprojection pipeline to mitigate the systematics induced by beam mismatches. Finally, we summarize the simulations used to assess the effectiveness of deprojection. In Sect.~\ref{sec:fg-cl-pip} we introduce the foreground cleaning pipelines including Needlet ILC (NILC) and constrained ILC (cILC), which we used to clean the $T$ or $E$ modes and $B$ modes, respectively. We present the results of the mock data after deprojection and foreground cleaning, carry out a lensing reconstruction process on the foreground-cleaned maps, {and constrain the tensor-to-scalar ratio $r$ using the cleaned $B$-mode power spectrum} in Sect.~\ref{sec:res}. Finally, we conclude in Sect.~\ref{sec:concl}.

\section{Beam systematics and deprojection}
\label{sec:bs}
The beam is the response of an antenna to sky radiation as a function of angle. Basically, each pixel on the focal plane includes a pair of two orthogonally polarized antennas that observe the same direction on the sky. The {time streams} from detectors \( a \) and \( b \) (a pair) are summed to obtain the total intensity (temperature) measurement, and they are differenced to cancel the unpolarized component. The radiation polarization is measured in this way. The typical beams of S3 are circular Gaussians with FWHM = 19 arcmin for 95 GHz, and 11 arcmin for 150 GHz, where there is no systematics in the ideal case. However, when the beams of the two antennas are mismatched, which means that their instrumental responses are different, the temperature signal is not completely canceled and could induce a spurious polarization signal in the pair difference data of the detector pair.\footnote{If the beams of the two detectors within a pair were distorted identically, the total miscalibration effect would not induce the $T\rightarrow P$ leakage, but other systematics instead \citep{collaborationBICEPKeckXVII2023}. Several filtering systematics that may cause the $E\rightarrow B$ leakage have been discussed in \citet{ghoshPerformanceForecastsPrimordial2022}.} 
For the $i$th detector, the TOD stream reads
\begin{equation}\label{eq:tod}
\begin{aligned}
    \tau_i(t)&=g_i\int d\nu\int d\Omega' A_i(\nu)B_i[\boldsymbol{\hat n}(t)-\boldsymbol{\hat n}']
    \\
    &[I'_{\nu}(\boldsymbol{\hat n}')+Q'_{\nu}(\boldsymbol{\hat n}')\cos2\psi_i(t)+U'_{\nu}(\boldsymbol{\hat n}')\sin2\psi_i(t)]\;.
\end{aligned}
\end{equation}
Of these, \( g_i \) is the gain of the detector, \( A_i(\nu) \) is the response of this detector at different frequencies, and \( B_i[\boldsymbol{\hat n}(t) - \boldsymbol{\hat n}'] \) is the beam function, representing the spatial response of the detector. \( [I, Q, U](\boldsymbol{\hat n}, \nu) \) are the Stokes parameters of the cosmic microwave background radiation at frequency \( \nu \) and position \( \boldsymbol{\hat n} \). The terms \( \boldsymbol{\hat n}(t) \) and \( \psi_i(t) \) are determined by the scan strategy. In an ideal case, \( B(\boldsymbol{\hat n}) \) and \( A_i(\nu) \) are delta functions. The beam systematic errors are included in the following way: 

\begin{enumerate}

\item The pointing error refers to the deviation between the pointing direction read by the instrument \( \boldsymbol{\hat n}_{\text{Ins}} \) and the true pointing direction of the telescope \( \boldsymbol{\hat n} \).
\item The band pass mismatch for a pair of detectors \( a \) and \( b \) means \( A_a(\nu) \neq A_b(\nu) \), which can introduce false signals when taking the difference between the readings of the detectors.
\item The beam mismatch for a pair of detectors \( a \) and \( b \) similarly refers to \( B_a(r) \neq B_b(r) \), which can also introduce false signals in the detector difference.
\item The gain mismatch arises when \( g_a \neq g_b \). Like the previous two points, this can introduce false signals in the detector difference.
\item The detector polarization angle calibration is typically detected by two detectors set to an angle of $90^{\circ}$. However, due to the precision of the detector manufacturing, the preferred polarization angle can deviate, that is, \( \psi_a(t) - \psi_b(t) \neq 90^\circ \), which can introduce systematic errors during the subsequent map-making process. 
\end{enumerate}
The first four errors above are all equivalent to an overall miscalibration of the detectors and a differential response between a pair of detectors at the same frequency, where we refer to the latter as beam mismatch in the following. A detector polarization angle miscalibration would introduce the $E\rightarrow B$ leakage, which is beyond the scope of this work. Given the high amplitude of the CMB temperature anisotropies relative to the polarization signal, the leakage from $T$ to $P$ must be carefully taken into account. The $T\rightarrow P$ leakage due to beam mismatches can be written as the convolution of the unpolarized signal and the differential beam within a detector pair,
\begin{equation}\label{eq:dtp}
    d_{T\rightarrow P}=T(\boldsymbol{\hat n})*[B_a(\boldsymbol{\hat n})-B_b(\boldsymbol{\hat n})]=T(\boldsymbol{\hat n})*B_\delta(\boldsymbol{\hat n})\,,
\end{equation}
where $T(\boldsymbol{\hat n})$ refers to the temperature anisotropies, $B_a(\boldsymbol{\hat n})$ and $B_b(\boldsymbol{\hat n})$ are the beams of two detectors, and $B_\delta(\boldsymbol{\hat n})$ is the differential beam of the pair.

\begin{table*}[htbp]
    \centering
    \caption{Six beam mismatch parameters added to the mock {TOD} and the corresponding leakage templates.}
    \renewcommand\arraystretch{1.3}
    \begin{tabular}{lccccccc}
        \bottomrule
        \bottomrule
        Mode&Gain&Beamwidth&Pointing, x&Pointing, y&Ellipticity, +&Ellipticity, $\times$ \\
        \hline
        Symbol&$\delta g$&$\delta \sigma$&$\delta x$&$\delta y$&$\delta p$&$\delta c$ \\
        \hline        
        Template\tablefootmark{a}&$\tilde T$&$(\nabla^2_x+\nabla^2_y) \tilde T$&$\nabla_x \tilde T$&$\nabla_y \tilde T$&$(\nabla^2_x-\nabla^2_y) \tilde T$&$2\nabla_x\nabla_y \tilde T$\\
        \hline
        95GHz mean &$3.9\times10^{-4}$&$-0.0026'$&$-0.9'$&$0.72'$&$4.3\times10^{-3}$&$-8.9\times10^{-3}$\\ 
        \hline
        95GHz std &$3\times10^{-3}$&$0.072'$&$0.35'$&$0.29'$&$1.4\times10^{-2}$&$1.3\times10^{-2}$\\ 
        \hline
        150GHz mean &$2.7\times10^{-4}$&$-0.0043'$&$-0.9'$&$0.84'$&$2.5\times10^{-3}$&$-3.2\times10^{-3}$\\ 
        \hline
        150GHz std &$3\times10^{-3}$&$0.072'$&$0.39'$&$0.35'$&$1.5\times10^{-2}$&$1.5\times10^{-2}$\\ 
        \hline
        \toprule
    \end{tabular}
    \\
    \tablefoottext{a}{$\tilde T$ is the temperature field convolved by the nominal beam. All six templates are $\tilde T$ and its first and second derivatives.}
    \label{tab:depr_temp}
\end{table*}


As in \citet{bicep2collaborationBICEP2IIIInstrumental2015} (hereafter BKIII), we modeled $B_\delta(\boldsymbol{\hat n})$ as the difference of two elliptical Gaussian beams for simplicity. This can be parameterized by six parameters (or differential modes): the gain difference $\delta g$, the differences of the pointing center coordinates (with respect to the nominal beam center) $\delta x$ and $\delta y$, the Gaussian beam-width difference $\delta\sigma$, and the differences of the plus- and cross-ellipticity $\delta p$ and $\delta c$. {The plus-ellipticity describes an ellipse whose major axis is oriented horizontally or vertically, and the cross-ellipticity describes an ellipse whose major axis is oriented $\pm45^\circ$ with respect to the x-axis} (see more details in BKIII, Appendix~A). For each differential mode $k=\{g,x,y,\sigma,p,c\}$, $\delta k=k_a-k_b$ is the difference of the beam parameter between two detectors. $B_\delta(\boldsymbol{\hat n})$ is the linear combination of the beams of the six modes, where each mode independently produces a pattern of the spurious polarized signal, also called a leakage template. For instance, since the beam of differential gain $B_{\delta g}(\boldsymbol{\hat n})$ is simply the scaled version of the circular Gaussian beam, its leakage is the scaled version of the beam-smoothed temperature $\tilde T(\boldsymbol{\hat n})$ according to Eq.~(\ref{eq:dtp}). Under the assumption that the mismatch is small, except {the third and higher orders}, all the templates are the beam-smoothed map of $T(\boldsymbol{\hat n})$ and its first and second spatial derivatives (see Table~\ref{tab:depr_temp}) \citep{Hu:2003}. The map-making procedure is a linear operation, meaning that the total leakage is the linear combination of the leakage templates of these differential modes. Therefore, we filtered the leakage out by fitting these templates to our data and then subtracting them.

The leakage templates we used are from the Planck 100 and 143 GHz reobserved TOD, which we used to clean 95 and 150 GHz, respectively. The reobserved templates were produced as follows: 
\begin{enumerate}
    \item We first generated CMB, foreground, and noise simulations of the Planck HFI bands as introduced below.
    \item {We coadded all components and reconvolved the coadded sky maps with the S3 nominal beams. For Planck 100 GHz, we divided $a_{\ell m}$ by the 100 GHz beam function and then multiplied them by the S3 95 GHz Gaussian beam. (This can be done on the input maps since the Planck and S3 nominal beams are circularly symmetric.)}
    \item {We computed the first and second derivatives of the Planck $T$ maps.}\footnote{{Although the deprojection technique is implemented on time streams of partial-sky observations, the computation of the first and second derivatives of $T$ maps is performed on Planck full-sky simulated data, thus being immune to the patch’s borders effects.}}
    \item {We rescanned the $T$ maps and their derivatives for each HFI band using the same pointing matrix as in S3.}
\end{enumerate}
{For the realistic observations, we reobserved the PR4 (NPIPE) maps \citep{planckcollaborationPlanckIntermediateResults2020} using the S3 scan strategy.} Our templates contained the noise component of the Planck HFI bands, which is not involved in the true $T\rightarrow P$ leakage, but we assumed that the noise is {much} lower than the temperature signal. {We produced simulated S3 TOD from the CMB-plus-foreground coadded maps using Eq.~(\ref{eq:tod}), and we computed the pair-difference time stream of the TOD by differencing the \( a \) and \( b \) detector time streams within a pair: $\tau_{\mathrm diff}(t)=(\tau_a-\tau_b)(t)/2$.} We finally fit the templates to the pair difference TOD and subtracted the fitted templates. Noise is not involved in the data since we expect that the noise is uncorrelated to the templates from Planck HFI maps, and the separation of noise ensured clearer analyses of the different components in the following foreground-cleaning pipeline.



{The assumed S3 experiment is located in the northern hemisphere. In its first observing season, it will scan 17\% of the sky where the thermal dust emission is clean, centering at $\mathrm{RA}=170^\circ$ and $\mathrm{DEC}=40^\circ$ in the celestial coordinate system. The polarized maps have a map depth of about 18 $\mu$K-arcmin at 95 GHz and 24 $\mu$K-arcmin at 150 GHz.}

Our mock dataset consisted of 300 simulated sky maps at $N_{\mathrm side}=1024$ in seven frequency bands, including the 95 and 150 GHz dual bands, the Planck HFI 100, 143, 217, 353 GHz bands, and the WMAP K band. The additional Planck and WMAP channels were for the purpose of cleaning the foregrounds more thoroughly. Each of the sky maps contained CMB, foregrounds, and white noise. First, we drew the six $\Lambda$CDM cosmological parameters within one standard deviation of the Planck 2018 best-fit cosmological constraints \citep{aghanimPlanck2018Results2020a}, from which we generated the lensed CMB power spectrum by \texttt{CAMB} \citep{lewisEfficientComputationCMB2000}. The CMB realizations were generated from the lensed power spectrum with the tensor-to-scalar ratio $r=0$, using the \textbf{synfast} facility from \texttt{lenspyx} \citep{reineckeImprovedCMBLensing2023}. The Galactic foregrounds containing thermal dust, synchrotron, free-free, spinning dust, and CO emissions were produced using the Planck Sky Model (\texttt{PSM}) package \citep{delabrouillePrelaunchPlanckSky2013}, which is the same as the foregrounds used in previous papers \citep{douForegroundRemovalILC2024,Han:2023gvr}. {The thermal dust polarization maps were generated by scaling the Planck 2018 GNILC polarized-dust template \citep{planckcollaborationPlanck2018Results2020a} to different frequencies using a modified blackbody SED with the dust temperature and spectral indices adopted from the best fit of the Planck 2015 GNILC dust maps \citep{planckcollaborationPlanckIntermediateResults2016}. The synchrotron polarization template was based on the Planck 2018 SMICA map and was scaled by a power-law SED with a fixed $\beta_s$ of -3.08. We did not use the current best-fit values to avoid confirmation biases during foreground removal.} For the 95 and 150 GHz bands alone, {we generated the TOD simulations of CMB plus foregrounds following Eq.~(\ref{eq:tod}) and the S3 scan strategy,} where we assumed the stochastic beam mismatch parameters whose mean and uncertainty are listed in Table~\ref{tab:depr_temp}. {The values were chosen to be at the same level as the measured beam parameters of the BICEP2 instruments \citep{adeBICEP2IIExperiment2014,bicep2BICEP2KeckArray2015}. They measured the beam shape parameters for each detector using a chopped thermal source mounted on a mast, and calibrated the differential gains using the cross-correlation of temperature maps for individual detectors with Planck.} We created white-noise simulations assuming that the realistic $1/f$ noise can be significantly removed by pair differencing and TOD polynomial filtering \citep{adeBICEP2IIExperiment2014}. The {white-noise simulations unrelated to systematics} were generated from the noise covariance matrices of the S3 and WMAP-K bands \citep{bennettNineYearWilkinsonMicrowave2013}. For the Planck HFI bands, we adopted 300 FFP10 noise simulations from the Planck Legacy Archive\footnote{http://pla.esac.esa.int/pla}. The deprojection process was performed on the CMB-plus-foreground data assuming that the effects on the noise component are negligible. 

\section{Foreground-cleaning pipeline}
\label{sec:fg-cl-pip}
We implemented the same foreground-cleaning methods as \citet{Han:2023gvr} to qualify the deprojection performance on the foreground-cleaned maps. The $T$ and $E$ modes were cleaned by the NILC method, and the $B$ modes were cleaned by the cILC method because the foreground dominates the $B$-mode signal.

The maps after foreground cleaning (NILC or cILC) consisted of four ingredients: the CMB signal, the residual foregrounds, the residual noise, and the systematic residual. For the purpose of examining the efficacy of the foreground cleaning methods, each component in the output map can theoretically be obtained by projecting ILC weights on the maps of the input component, given the linearity of ILC methods. However, the input CMB and foregrounds are mixed up during deprojection so that the systematic contamination on the CMB and foregrounds can never be separated afterward. {Instead}, we {first} subtracted the residual noise from the total residual (the foreground-cleaned map minus the input CMB map), which was the sum of the residual foregrounds and the systematic residual. {We then projected the ILC weights on the deprojection residual maps at the 95 and 150 GHz bands to obtain the deprojection residual in the foreground-cleaned maps. Finally, we subtracted it from the sum to compute the residual foregrounds.}

\subsection{The NILC pipeline}
\label{sec:nilc-pip}
The needlet internal linear combination (NILC) method \citep{delabrouilleFullSkyLow2009,basakNeedletILCAnalysis2012,basakNeedletILCAnalysis2013} is a widely used blind component-separation technique that linearly combines the raw data from different frequencies while minimizing the variance in the needlet space. Needlets are a special form of wavelets that permit a localization in harmonic and real {(pixel)} space. They are therefore suitable for component separation when the properties of the sky components vary significantly with the sky positions and scales. We adopted the NILC method for cleaning foregrounds in $T$- and $E$-mode maps, the details of which are described below.

The sky observations $d_\nu(p)$ at a frequency band $\nu$ and sky pixel $p$ were first transformed into spherical harmonic coefficients using \texttt{HEALPix}\footnote{http://healpix.sf.net/} \citep{gorskiHEALPixFrameworkHigh2005}: $d_{\ell m}^\nu=\int d_\nu(p)Y_{\ell m}^*(p)d\Omega$. The raw maps with different beams were then reconvolved into the common resolution by $d_{\ell m}^{\nu,\mathrm out}=d_{\ell m}^{\nu,\mathrm in}b_\ell^{\mathrm out}/b_\ell^{\mathrm in}$, where $b_\ell$ is the beam transfer function. The common beam resolution ($b_\ell^{\mathrm out}$) was chosen as a Gaussian beam with FWHM=11 arcmin, which is the resolution of the 150 GHz band. For the WMAP K band and 95 GHz, which have larger beam sizes than the common beam ($b_\ell^{\mathrm in}<b_\ell^{\mathrm out}$), we cut off the harmonic coefficients of these two bands with $\ell$ greater than 350 and 1200, respectively, to avoid an overamplification of the instrumental noise. The harmonic coefficients can be decomposed into a set of filtered maps,
\begin{equation}\label{eq:dlm_nu}
    d_{\ell m}^{\nu,j}=h_\ell^j d_{\ell m}^\nu\,,
\end{equation}
where the needlet bands $h_\ell^j$ satisfy $\sum_j(h_\ell^j)^2=1$, defining a localization scheme in harmonic space. We adopted the cosine needlet bands as shown in Fig.~1 of \citet{Han:2023gvr}. {The eight needlet bands peak at 15, 30, 60, 120, 300, 700, 1200, and 2000 with $\ell_{\max}=2000$.} For the WMAP K band, whose coefficients with $\ell>350$ were cut off, the needlets whose peaks ($\ell_{\mathrm peak}$) are greater than 350 would not be applied to the K-band data {(and so is for 95 GHz band)}. For each frequency band, the filtered maps were then transformed back to real space, forming a set of needlet maps.
The needlet coefficient of the $j$th needlet and the $k$th pixel after transformation is given by
\begin{equation}\label{eq:bj_nu}
    b^\nu_j(\hat p_{jk}) = \sqrt{\frac{4\pi}{N_j}}\sum_{\ell m} h^j_\ell d^\nu_{\ell m} Y_{\ell m}(\hat p_{jk})\,,
\end{equation}
where $N_j$ denotes the number of pixels of the $j$th needlet map, and $\hat p_{jk}$ denotes the pixel center of the $k$th pixel of the $j$th needlet map.

When the data {were} transformed to needlet space, we applied the internal linear combination method to the multichannel data. First we estimated the data covariance matrix across the frequencies by averaging the needlet coefficient product $b^{\nu_1}_j(\hat p_{jk})b^{\nu_2}_j(\hat p_{jk})$ over a disk of pixels centered at pixel $k$. The empirical data covariance between $\nu_1$ and $\nu_2$ is written as
\begin{equation}\label{eq:Cjk1x2}
    \hat C^{\nu_1\times \nu_2}_{jk} = \frac{1}{n_k} \sum_{k'} w_j(k, k') b^{\nu_1}_j(\hat p_{jk'}) b^{\nu_2}_j(\hat p_{jk'})\,,
\end{equation}
where $w_j(k, k')$ are the weights that select the region around a pixel $k$ to average out, and $n_k$ is the number of selected pixels. The NILC solution is a linear combination of the needlet coefficients of all frequencies,
\begin{equation}\label{eq:bj_nilc}
    b^{\mathrm NILC}_j(\hat p_{jk}) = \sum_{\nu} w^{\mathrm NILC}_{\nu, j}(\hat p_{jk}) b^\nu_j(\hat p_{jk})\,,
\end{equation}
where the NILC weights are given by
\begin{equation}
    w^{\mathrm NILC}_{\nu, j}(\hat p_{jk}) = \left[\frac{\boldsymbol {\hat C}_{jk}^{-1} \boldsymbol a}{\boldsymbol a^t\boldsymbol {\hat C}_{jk}^{-1} \boldsymbol a}\right]_\nu\,.
    \label{eq:nilc_wgts}
\end{equation}
Here, $\boldsymbol {\hat C}_{jk}$ is the $n_\nu\times n_\nu$ empirical data covariance matrix, where $n_\nu$ is the number of frequency channels, and $\boldsymbol a$ is the spectral response vector that characterizes the variation in the signal with frequencies. Because the CMB signal is frequency independent in thermodynamic temperature units, we set $\boldsymbol a$ as a unit column vector with a length of $n_\nu$. The relation $\sum_\nu a_\nu w^{\mathrm NILC}_{\nu, j}=1$ ensures the precise recovery of the CMB component.

Finally, we performed an inverse needlet transformation to obtain the NILC-cleaned map,
\begin{equation}\label{eq:slm_nilc}
    \hat s^{\mathrm NILC}_{\ell m} = \sum_{jk}b^{\mathrm NILC}_j(\hat p_{jk}) \sqrt{\frac{4\pi}{N_j}} h^j_\ell Y_{\ell m}(\hat p_{jk})\,.
\end{equation}
The recovered CMB map in real space is given by $\hat s_{\mathrm NILC} (p)= \sum_{\ell m} \hat s^{\mathrm NILC}_{\ell m} Y_{\ell m} (p)$.
{We used a mask that preserved the pixels with a noise variance of the 150 GHz band smaller than 20 $\mu$K pixels, whose sky fraction was about 10\% (see Fig.~\ref{fig:B150-maps}).}

\begin{figure*}[htbp]
    \centering
        \includegraphics[width=0.9\textwidth]{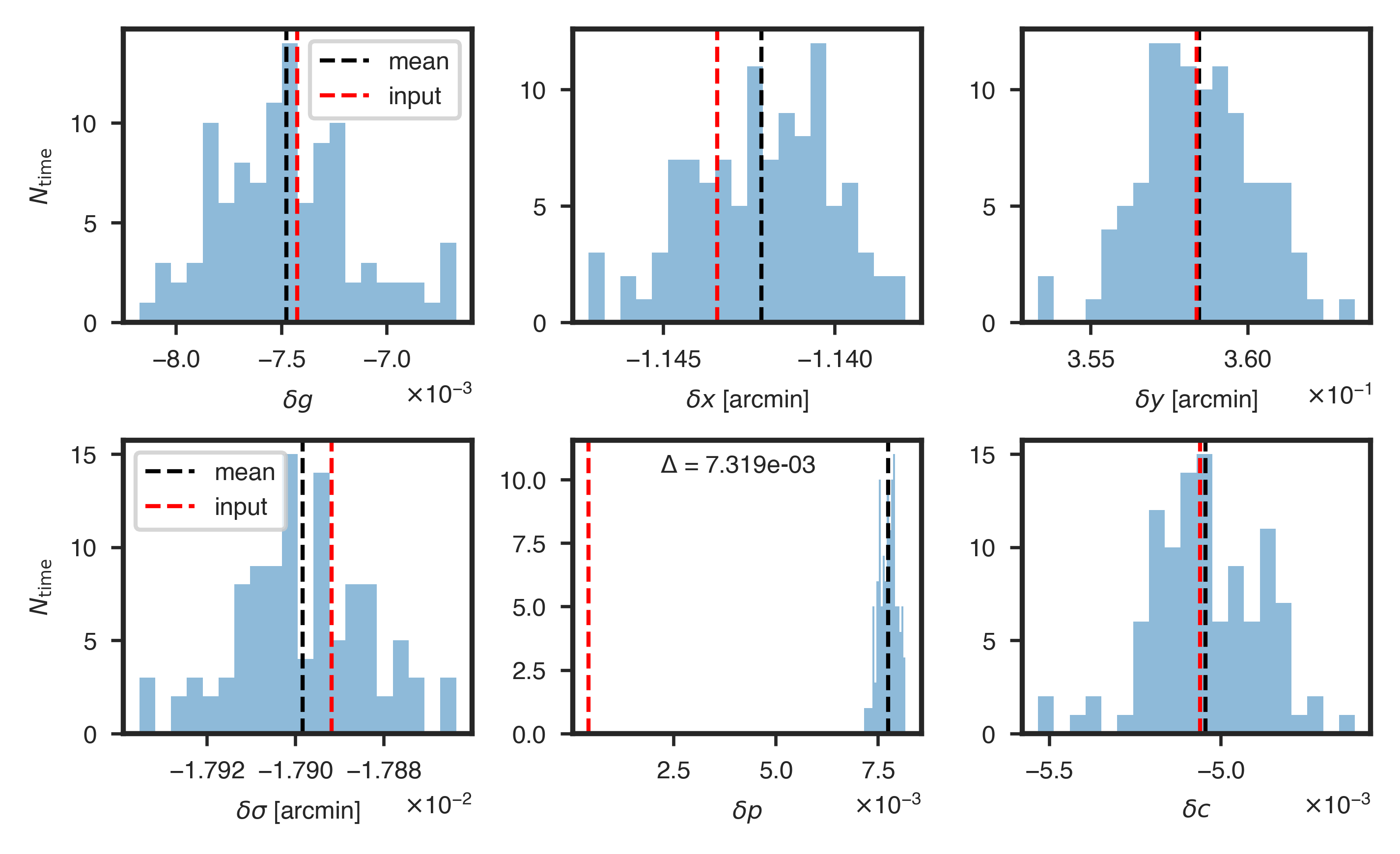}
    \caption{{Distribution of the recovered beam mismatch parameters of 106 time trunks for one detector pair of a noise-free simulation at 150 GHz. The dashed red lines represent the input beam parameter, and the dashed black lines indicate the mean over all time series. There is a bias on the differential plus-ellipticity of $\Delta\approx7.3\times10^{-3}$ due to the cosmological $TE$ correlation.}}
    \label{fig:bsp-150}
\end{figure*}

\subsection{The cILC pipeline}
\label{sec:cilc-pip}
Following the ILC method, the constrained ILC (cILC) method \citep{remazeillesCMBSZEffect2011,remazeillesPeelingForegroundsConstrained2021} was developed to further mitigate the residual foreground contaminants by adding constraints on the ILC weights. We used the cILC method in harmonic space to produce foreground-cleaned $B$-mode maps because the foregrounds are several orders of magnitude higher than the CMB signal for $B$ modes, and the standard ILC method is therefore insufficient to reduce the residual foregrounds to a level of $r=0.01$ primordial tensor modes under the sensitivities of S3.

We modeled the $B$-mode observation at a frequency $\nu$ and pixel $p$ as the sum of $n_c$ sky components and the thermal noise,
\begin{equation}\label{eq:dp_nu}
    d_\nu(p) = \sum_c A_{\nu c} s_c(p) + n_\nu(p)\,,
\end{equation}
where $s_c(p)$ is the astrophysical emission template of component $c$, and $n_\nu(p)$ is the noise. The $n_\nu\times n_c$ mixing matrix $\boldsymbol{A}$ stores the spectral response information that describes the SED of component $c$ at frequency $\nu$. In our implementation, we only considered three dominant sky components: the lensed CMB, the polarized thermal dust, and the polarized synchrotron. The first column of the mixing matrix refers to the mixing vector of CMB, $\boldsymbol a$. The second column corresponding to dust emissions was calculated assuming a modified blackbody SED with fixed $T_{\mathrm dust}=19.6K$ and $\beta_{\mathrm dust}=1.59$. The third column, that is, the mixing vector of the synchrotron, follows the power law SED with fixed $\beta_{\mathrm sync}=-3$ in Rayleigh-Jeans brightness temperature units.

The cILC weights are given by
\begin{equation}\label{eq:wgts_cilc}
    \boldsymbol{w}^{\mathrm cILC} = \boldsymbol{e} \left( \boldsymbol{A}^T\hat{\boldsymbol{C}}^{-1} \boldsymbol{A}\right)^{-1}\boldsymbol{A}^T \hat{\boldsymbol{C}}^{-1}\,,
\end{equation}
which satisfy the constraints of nulling the foreground components (dust and synchrotron),
\begin{equation}\label{eq:wAee}
    \sum_\nu w^{\mathrm cILC}_{\nu,\ell} A_{\nu,c} = e_c\,,
\end{equation}
where $\boldsymbol{e}$ is a row vector with length $n_c$, whose element $e_c$ is one for the CMB and zero for the foreground components. The empirical data covariance matrix $\hat{\boldsymbol{C}}$ is computed by
\begin{equation}\label{eq:Cl1x2}
    \hat C^{\nu_1\times \nu_2}_{\ell} = \frac{1}{\sum_{\ell'=\ell_{\min}}^{\ell_{\max}}(2\ell+1)}\sum_{\ell'=\ell_{\min}}^{\ell_{\max}}\sum_{m=-\ell'}^{\ell'} d_{\ell' m}^{\nu_1\,*} d_{\ell' m}^{\nu_2}\,,
\end{equation}
{where $\ell_{\min}=\min[0.6\ell, \ell-7]$, and $\ell_{\max}=\max[1.4\ell, \ell+7]$, and we excluded the center multipole ($\ell'\neq\ell$) to mitigate the ILC bias. The input maps were reconvolved to a common beam of 11 arcmin. The analysis was performed with $\ell_{\max}=2000$, while the harmonic coefficients of WMAP-K and 95 GHz bands were not involved as $\ell$ is greater than 350 and 1200, respectively.}
The cILC-cleaned CMB map was finally obtained by $\hat s_{\ell m}^{\mathrm cILC} = \sum_\nu w^{\mathrm cILC}_{\nu,\ell} d_{\nu,\ell m}$.
{To reduce the foreground contaminants, we used a smaller mask ($f_{\mathrm sky}\approx7\%$) to clean the $B$ maps, which was produced by removing the pixels with a noise variance in the 150 GHz band larger than 10 $\mu$K-pixel and a declination above 65$^\circ$ (see the third row of Fig.~\ref{fig:nilc-TEB-maps}).}

\begin{figure*}[htbp]
    \centering
    \begin{subfigure}[b]{0.3\textwidth}
        \centering
        \includegraphics[width=\textwidth]{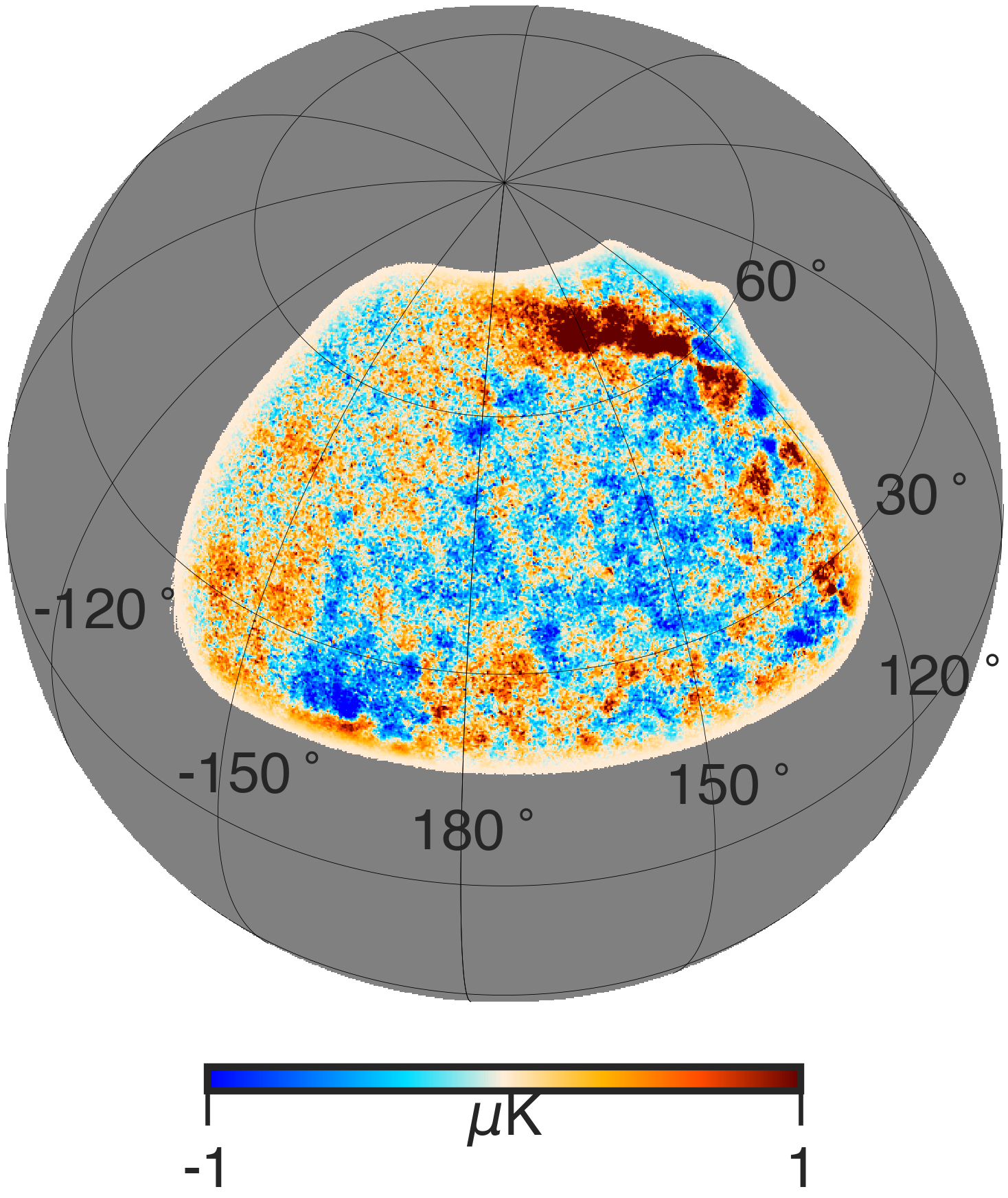}
        \caption{Input CMB+FG. {without beam systematics} of the $B$ modes}
    \end{subfigure}
    \begin{subfigure}[b]{0.3\textwidth}
        \centering
        \includegraphics[width=\textwidth]{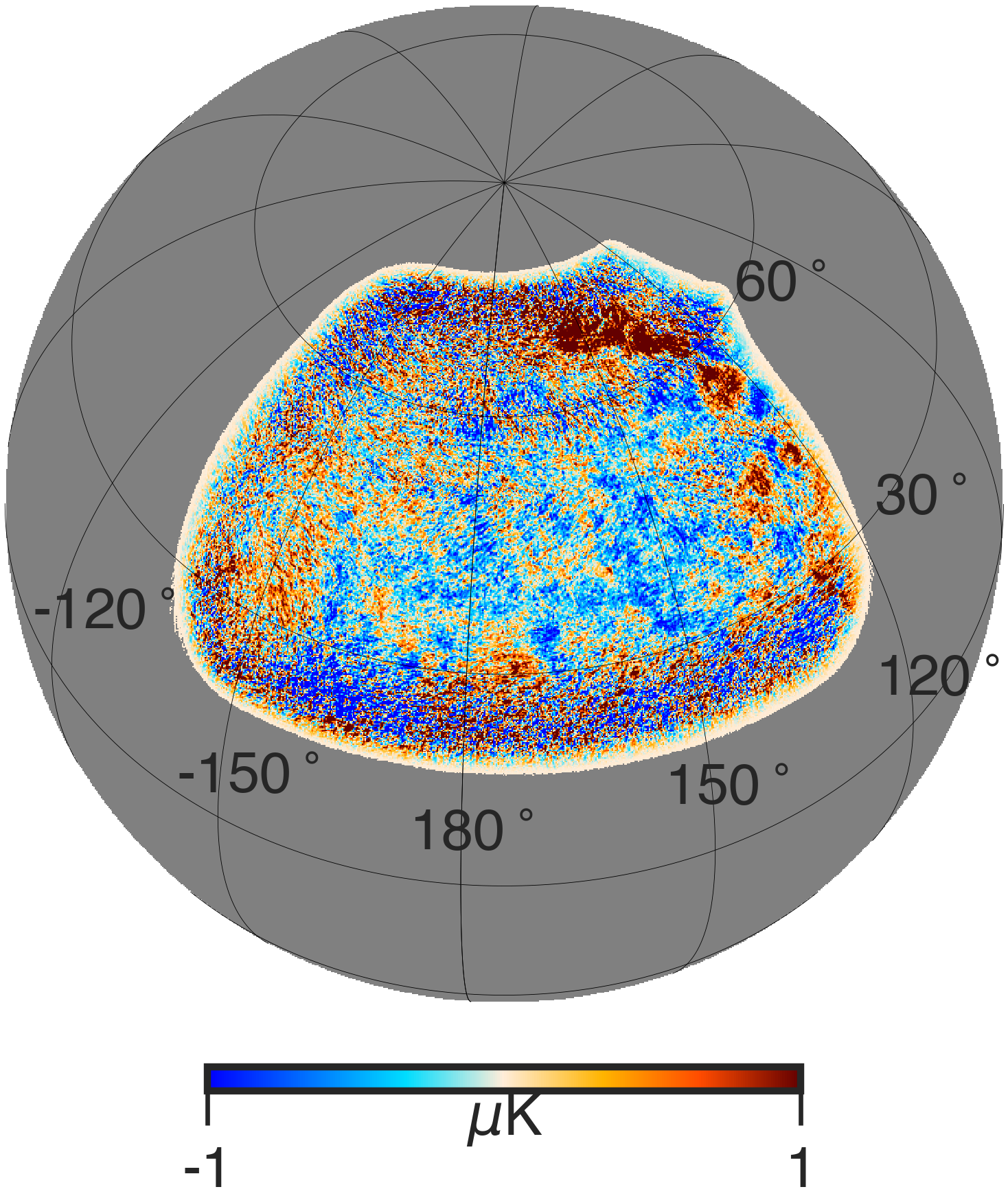}
        \caption{CMB+FG. with {beam systematics} of the $B$ modes}
    \end{subfigure}
    \begin{subfigure}[b]{0.3\textwidth}
        \centering
        \includegraphics[width=\textwidth]{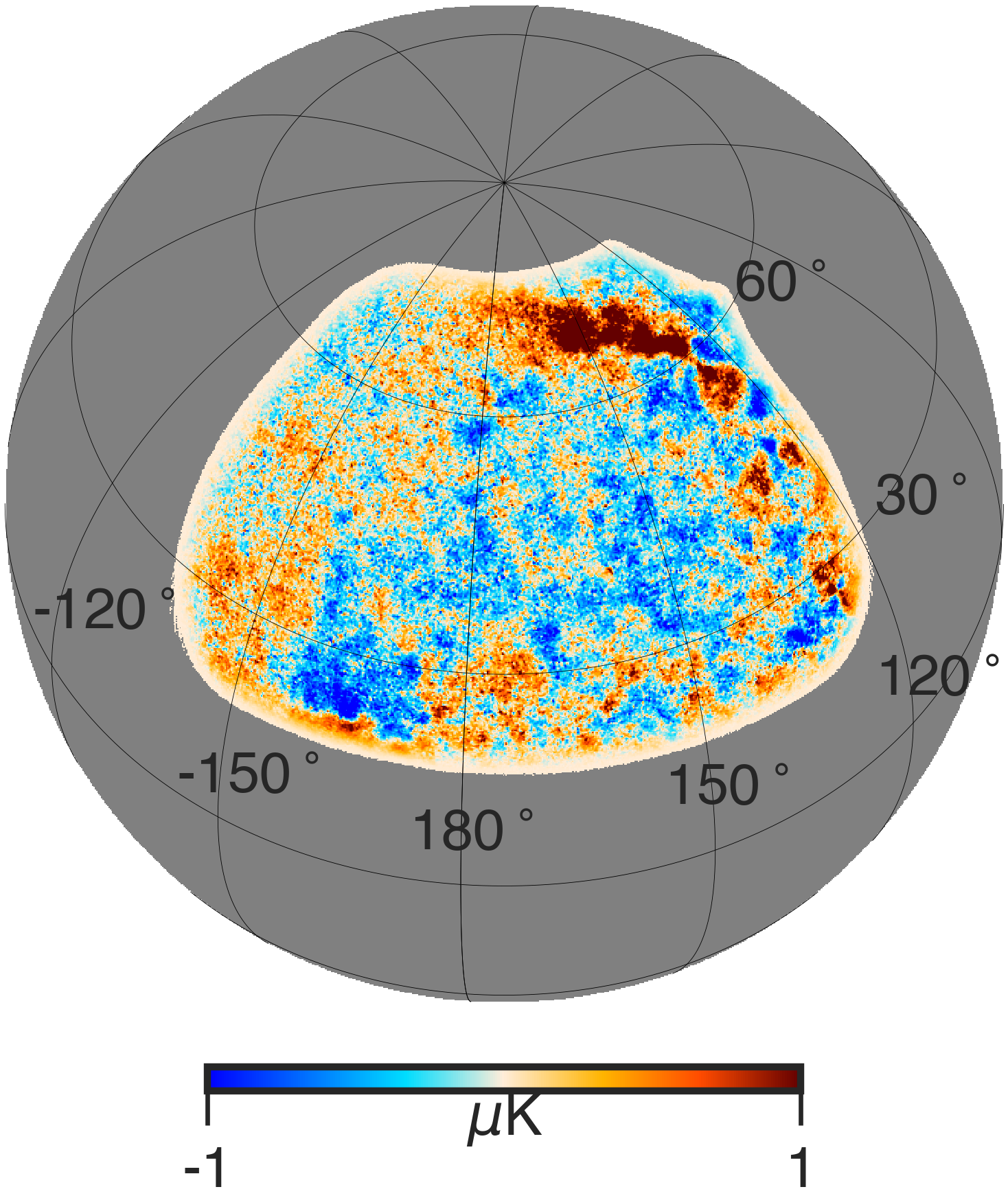}
        \caption{CMB+FG. after deprojection of the $B$ modes}
    \end{subfigure}
    \\
    \begin{subfigure}[b]{0.3\textwidth}
        \centering
        \includegraphics[width=\textwidth]{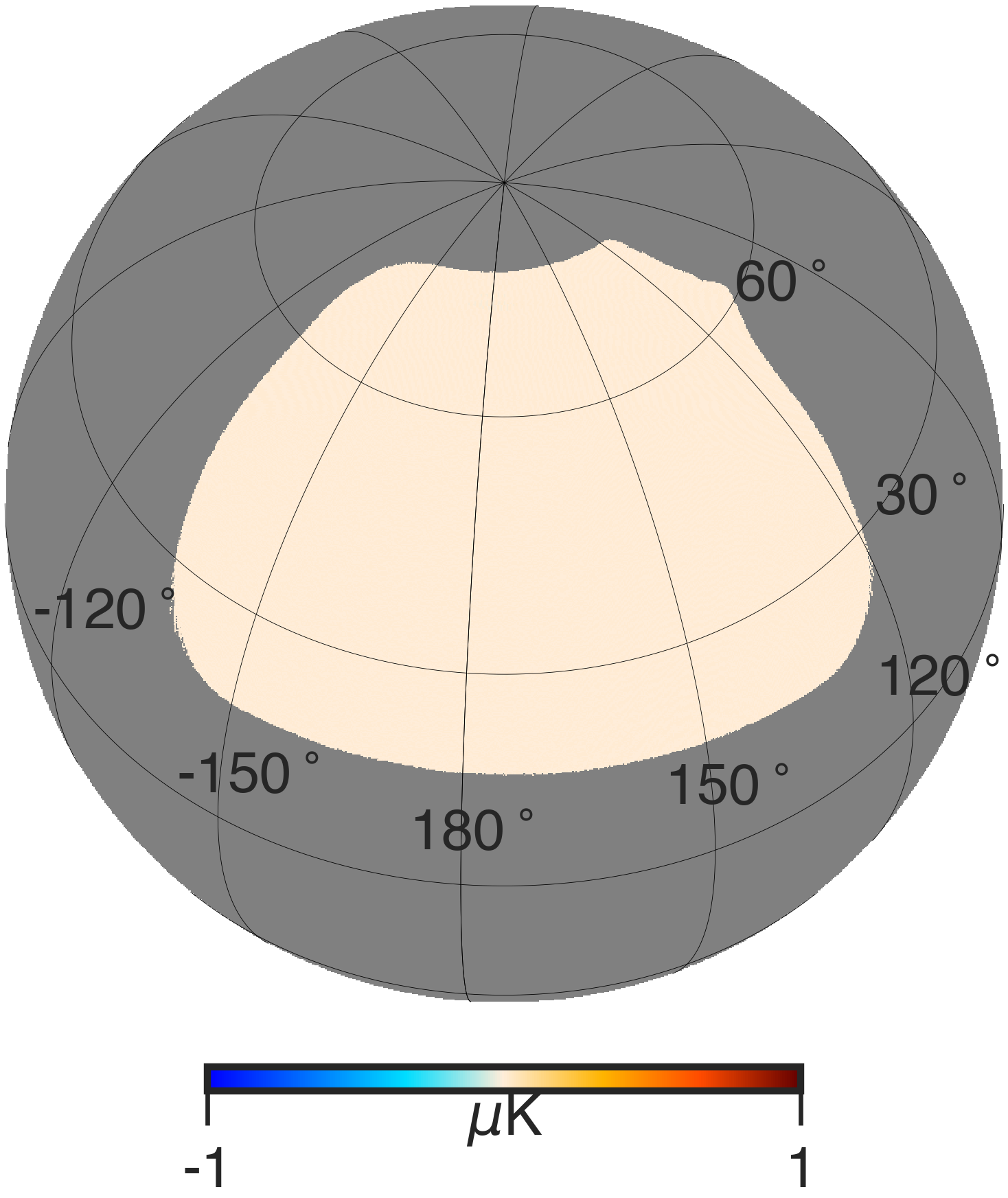}
        \caption{CMB+FG. deprojected residual of the $T$ modes}
    \end{subfigure}
    \begin{subfigure}[b]{0.3\textwidth}
        \centering
        \includegraphics[width=\textwidth]{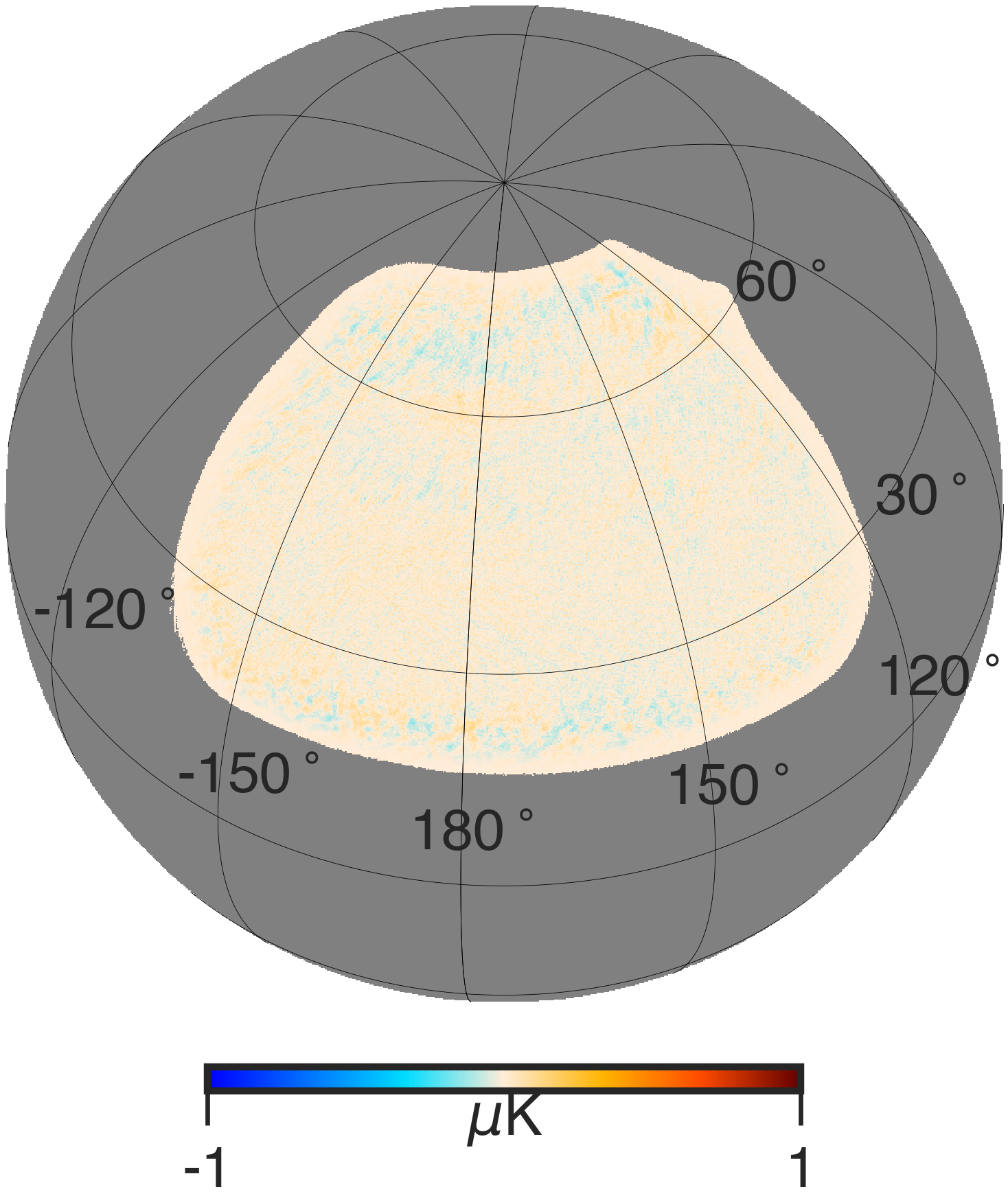}
        \caption{CMB+FG. deprojected residual of the $E$ modes}
    \end{subfigure}
    \begin{subfigure}[b]{0.3\textwidth}
        \centering
        \includegraphics[width=\textwidth]{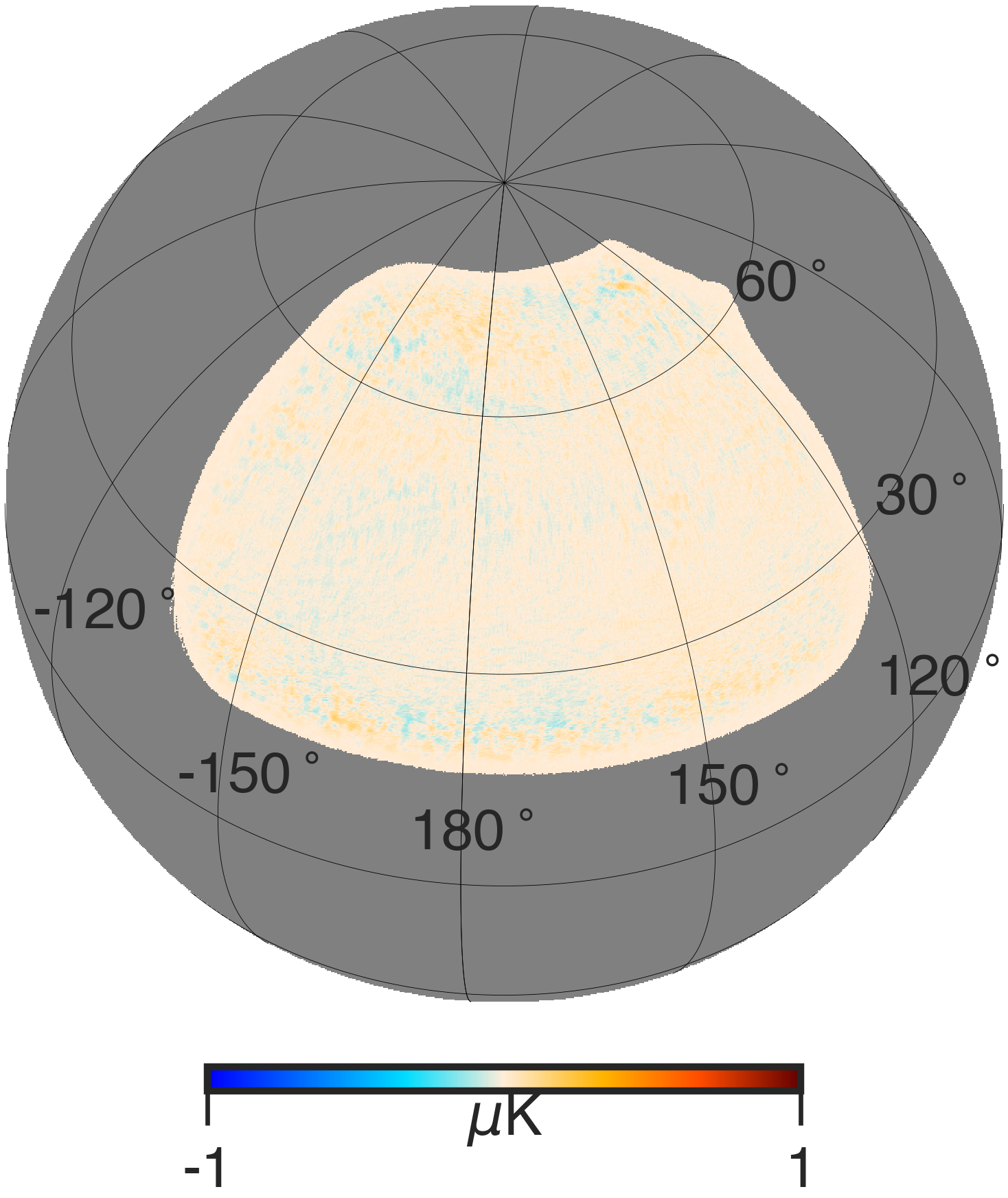}
        \caption{CMB+FG. deprojected residual of the $B$ modes}
    \end{subfigure}
    \caption{Maps of CMB plus foregrounds at S3 150 GHz. First row: Input $B$-mode CMB and foregrounds without beam systematics (left), CMB plus foregrounds with beam systematics (middle) and after deprojection (right). Second row: Difference between the deprojected map and the input map without beam systematics for the $T$ (left), $E$ (middle), and $B$ (right) modes. The results of 95 GHz are similar to the 150 GHz results.}
    \label{fig:B150-maps}
\end{figure*}

\begin{figure*}[htbp]
    \centering
        \includegraphics[width=0.9\textwidth]{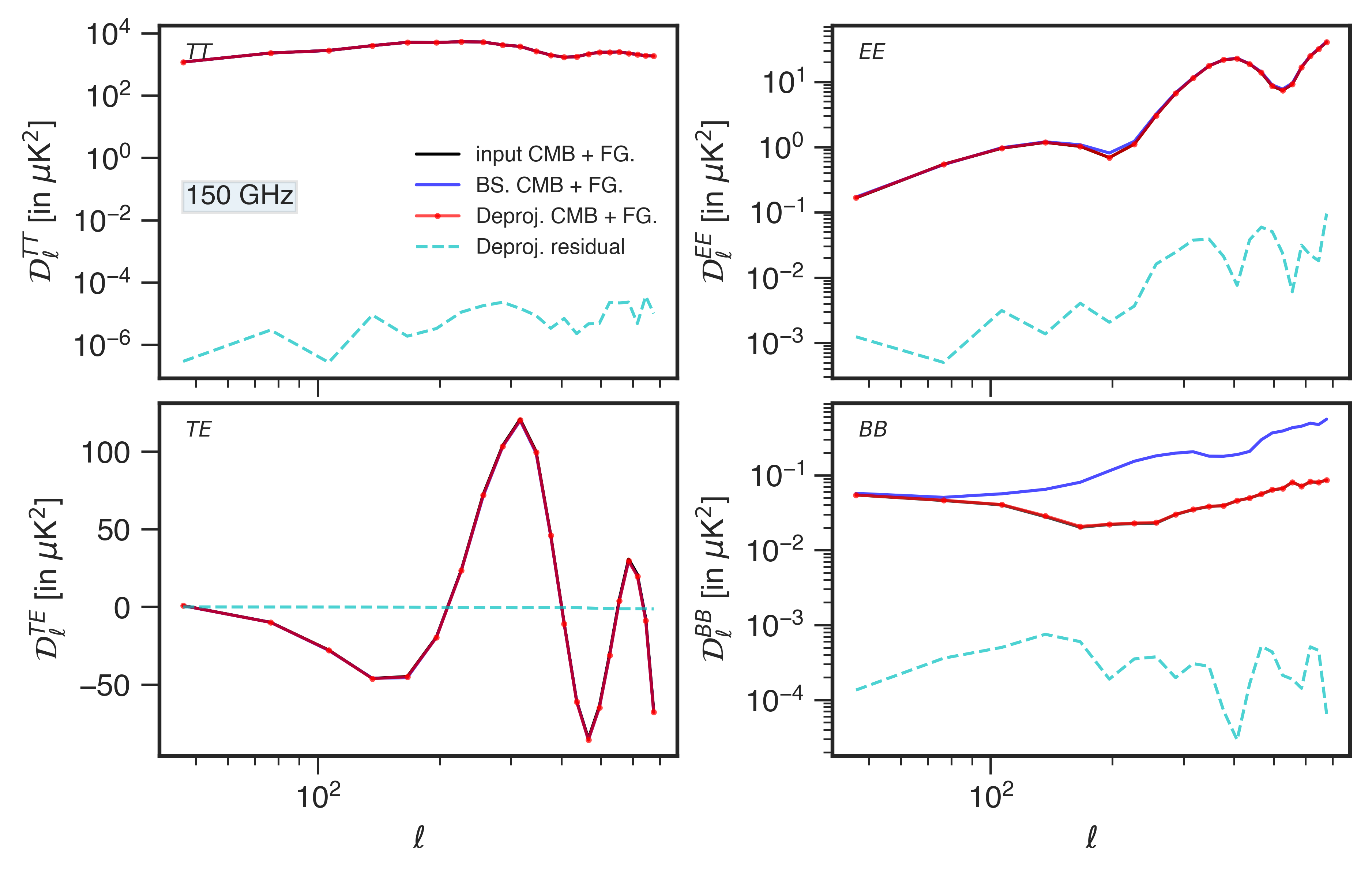}
    \caption{$TT$, $EE$, $TE$, and $BB$ power spectra of the 150 GHz CMB plus foregrounds map with beam systematics before (blue curves) and after deprojection (red curves) for one realization. The input CMB+FG. power spectrum without beam systematics is shown as a solid black curve for comparison with the deprojected one, {and their difference, the residual after deprojection, is shown as a dashed cyan curve.} The results of 95 GHz are similar to the 150 GHz results.}
    \label{fig:B150-ps}
\end{figure*}

\begin{figure*}[ht!]
    \centering
    \begin{subfigure}[b]{0.3\textwidth}
        \centering
        \includegraphics[width=\textwidth]{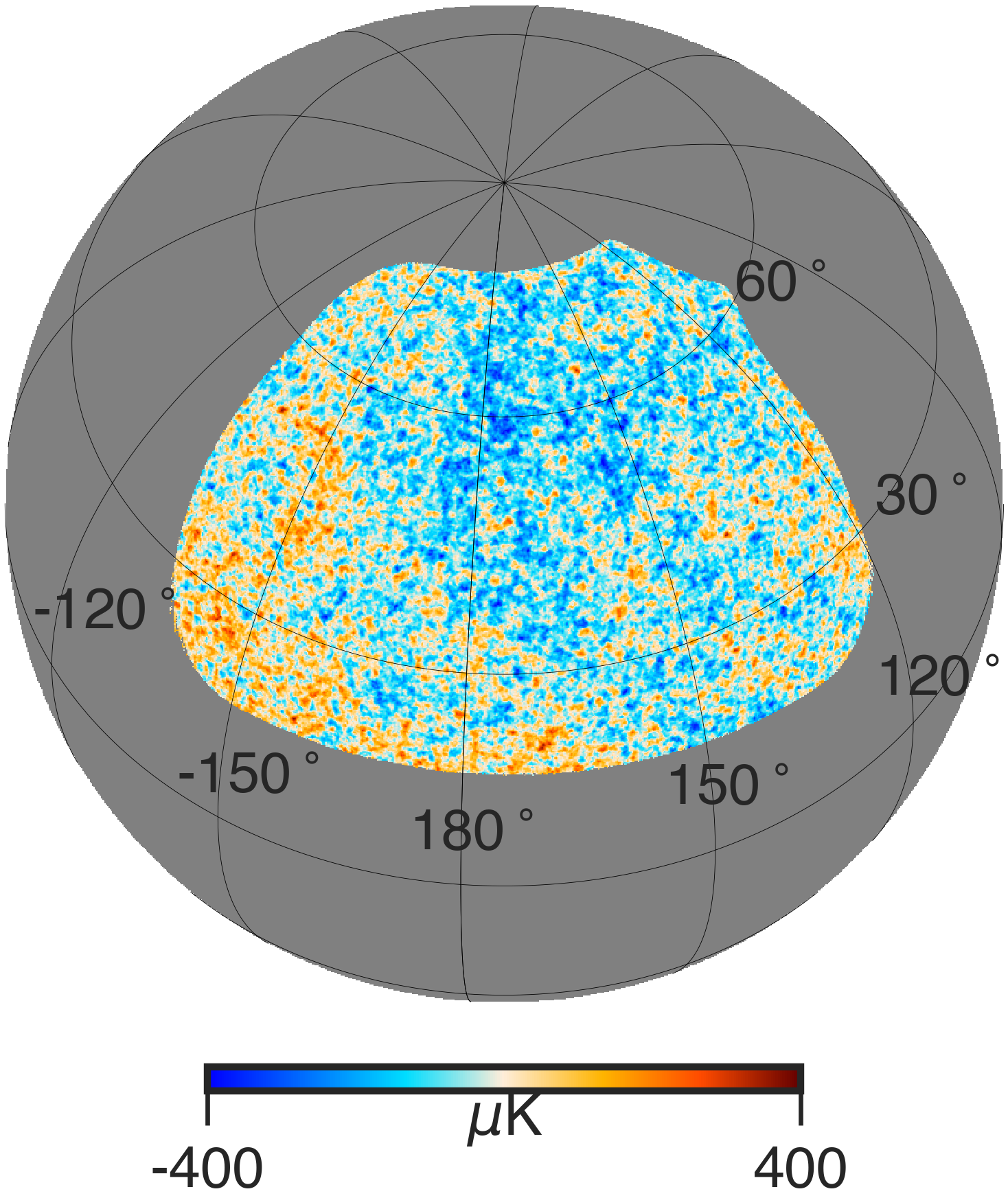}
        \caption{Input $T$ CMB}
    \end{subfigure}
    \begin{subfigure}[b]{0.3\textwidth}
        \centering
        \includegraphics[width=\textwidth]{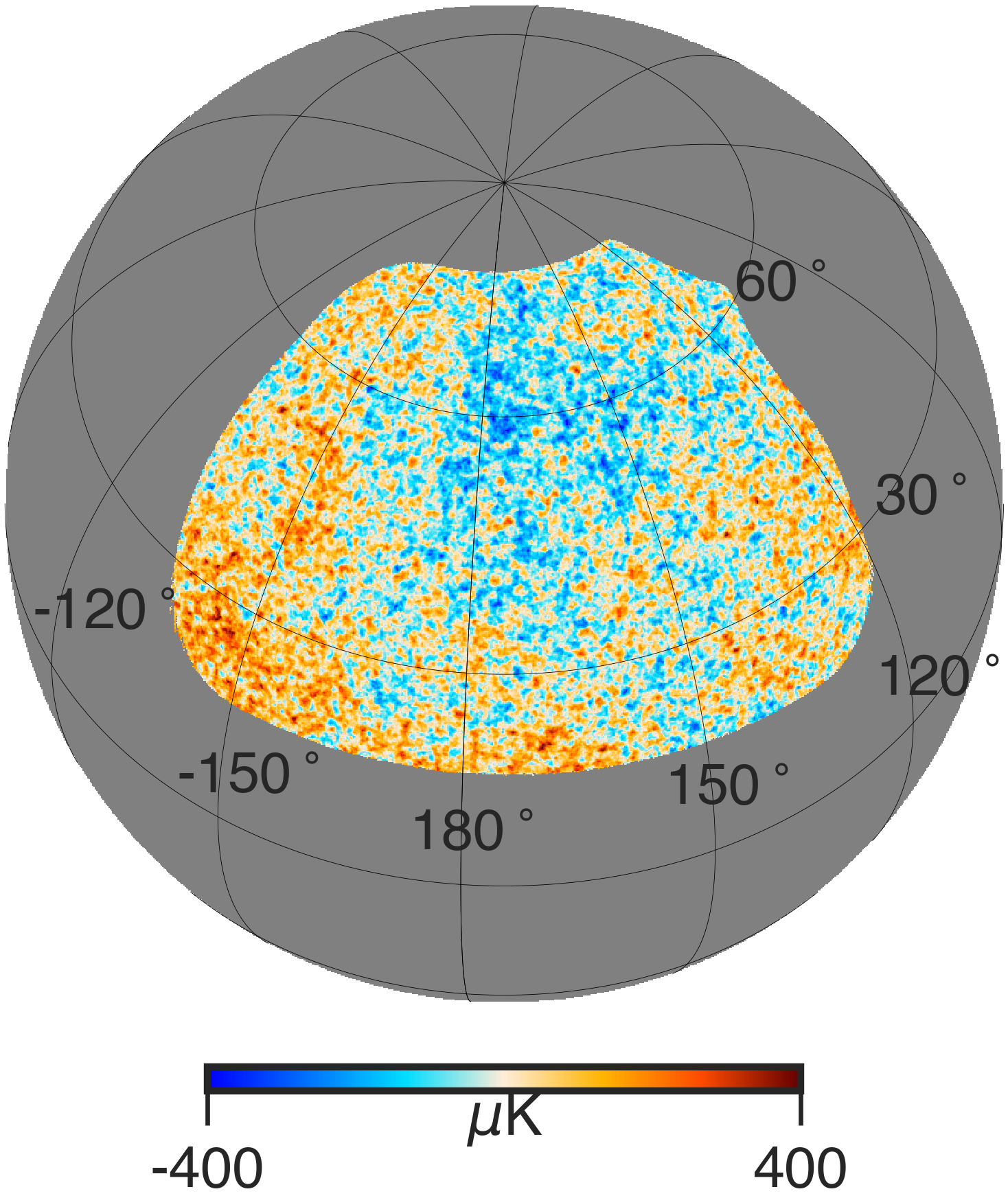}
        \caption{NILC-cleaned $T$ map}
    \end{subfigure}
    \begin{subfigure}[b]{0.3\textwidth}
        \centering
        \includegraphics[width=\textwidth]{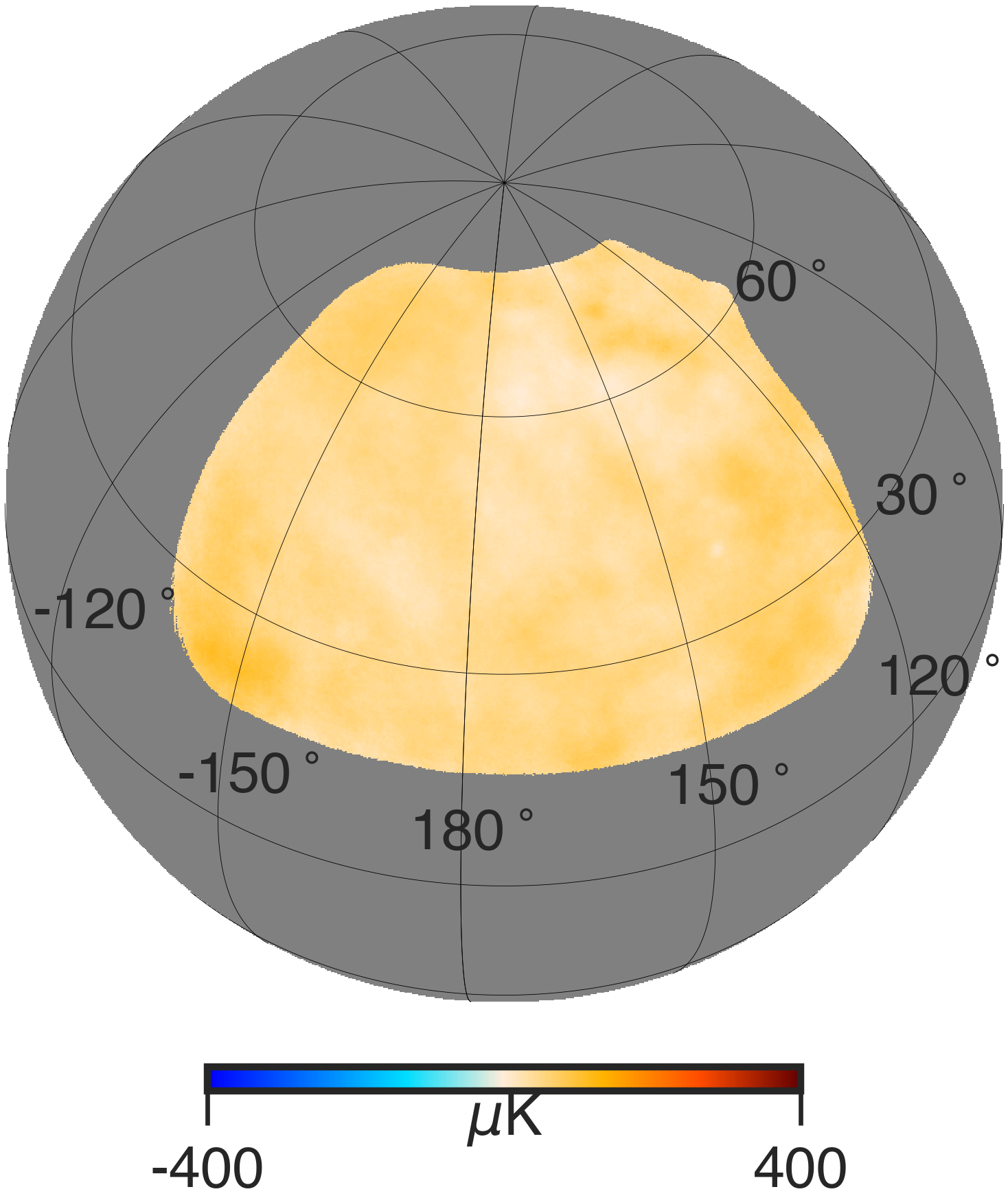}
        \caption{NILC residual $T$ map}
    \end{subfigure}
    \\
    \begin{subfigure}[b]{0.3\textwidth}
        \centering
        \includegraphics[width=\textwidth]{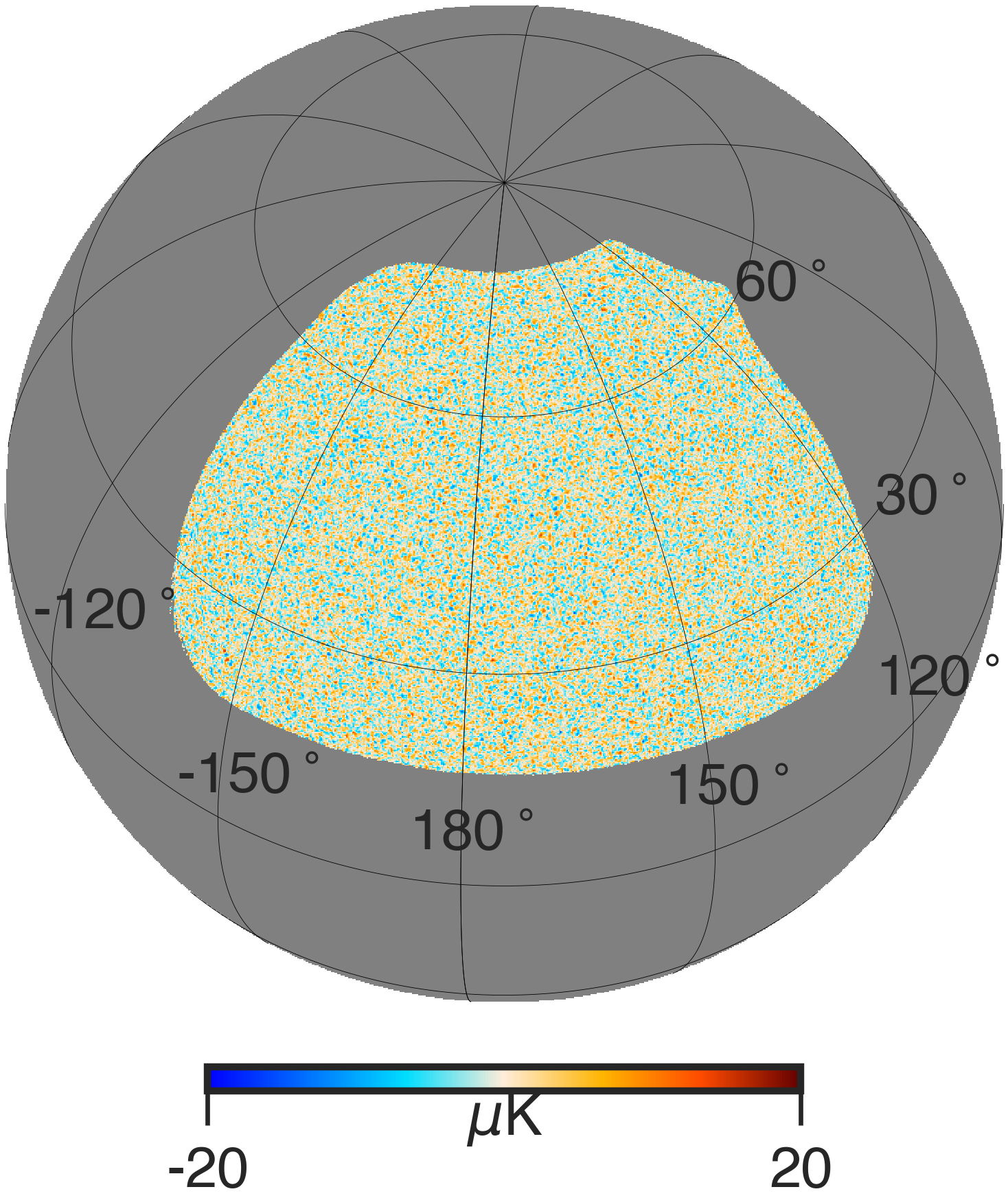}
        \caption{Input $E$ CMB}
    \end{subfigure}
    \begin{subfigure}[b]{0.3\textwidth}
        \centering
        \includegraphics[width=\textwidth]{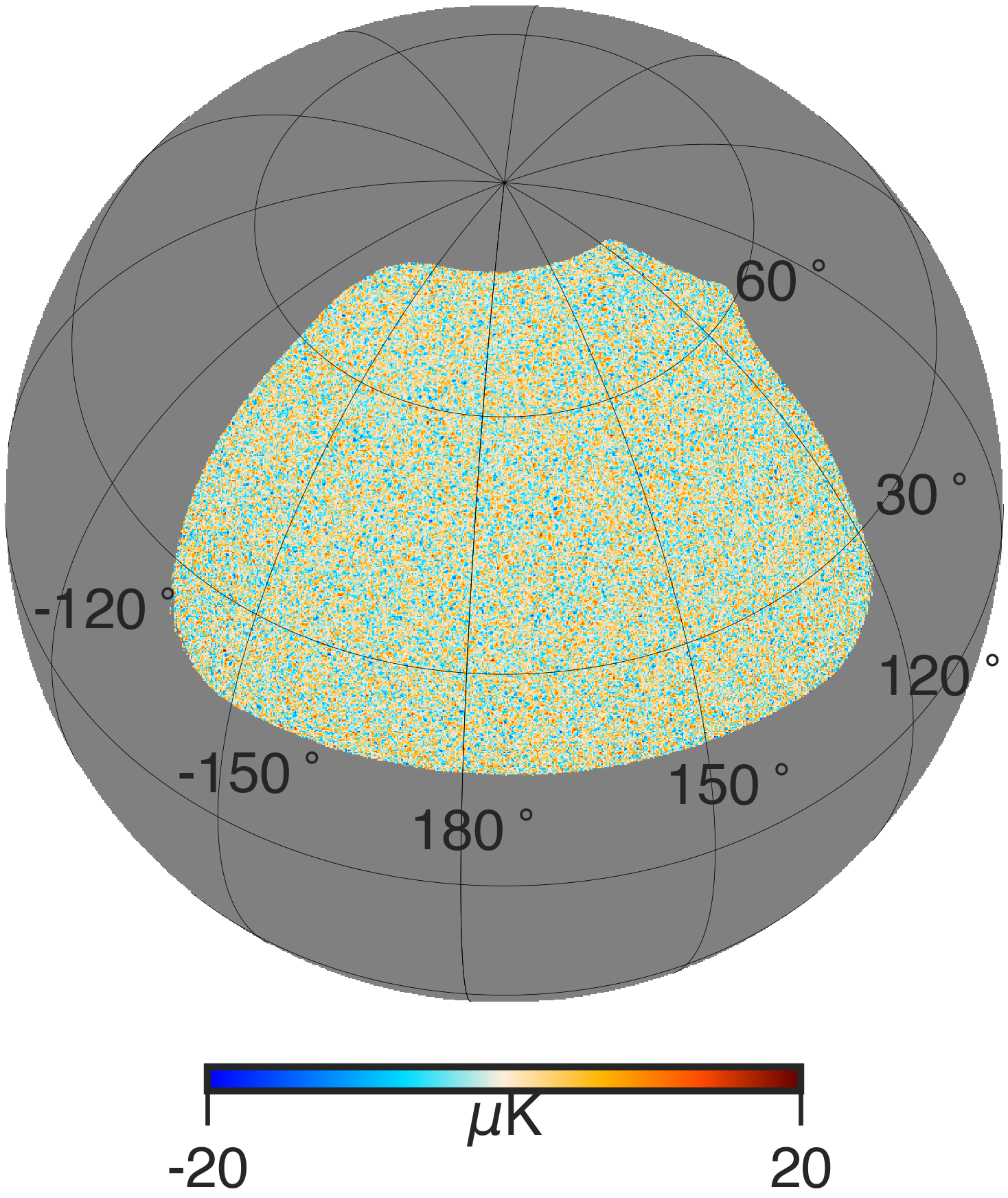}
        \caption{NILC-cleaned $E$ map}
    \end{subfigure}
    \begin{subfigure}[b]{0.3\textwidth}
        \centering
        \includegraphics[width=\textwidth]{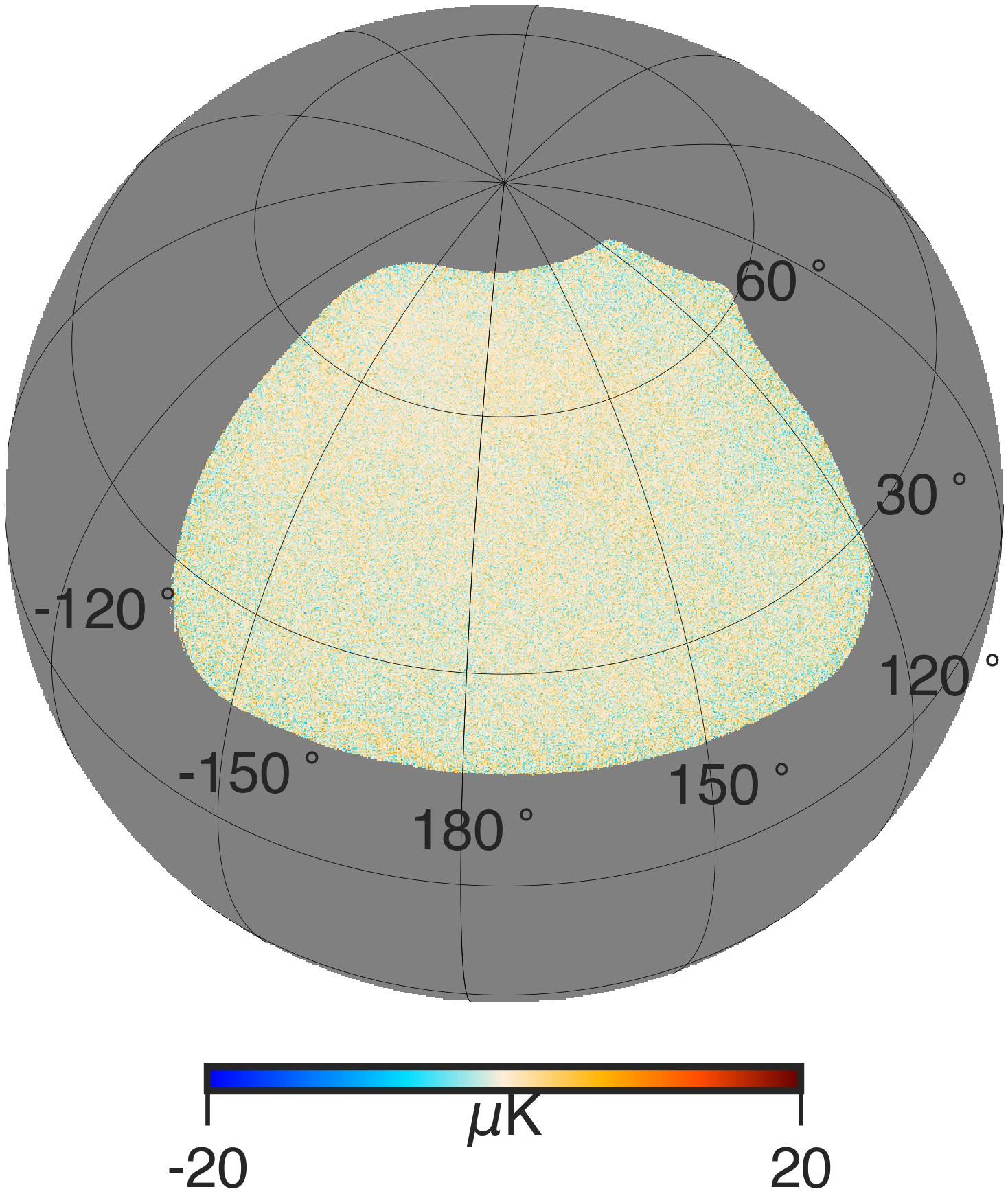}
        \caption{NILC residual $E$ map}
    \end{subfigure}
    \\
    \begin{subfigure}[b]{0.3\textwidth}
        \centering
        \includegraphics[width=\textwidth]{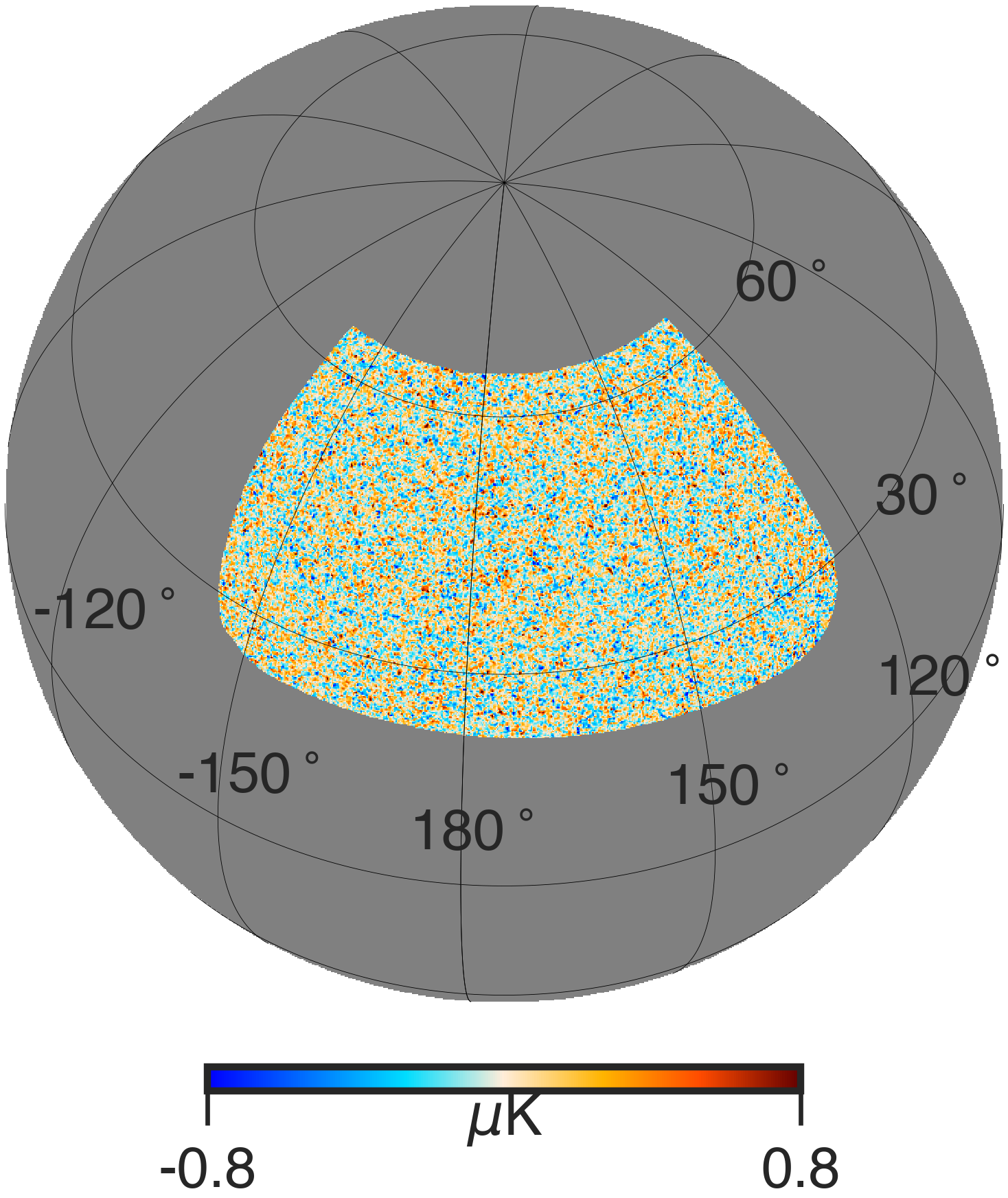}
        \caption{Input $B$ CMB}
    \end{subfigure}
    \begin{subfigure}[b]{0.3\textwidth}
        \centering
        \includegraphics[width=\textwidth]{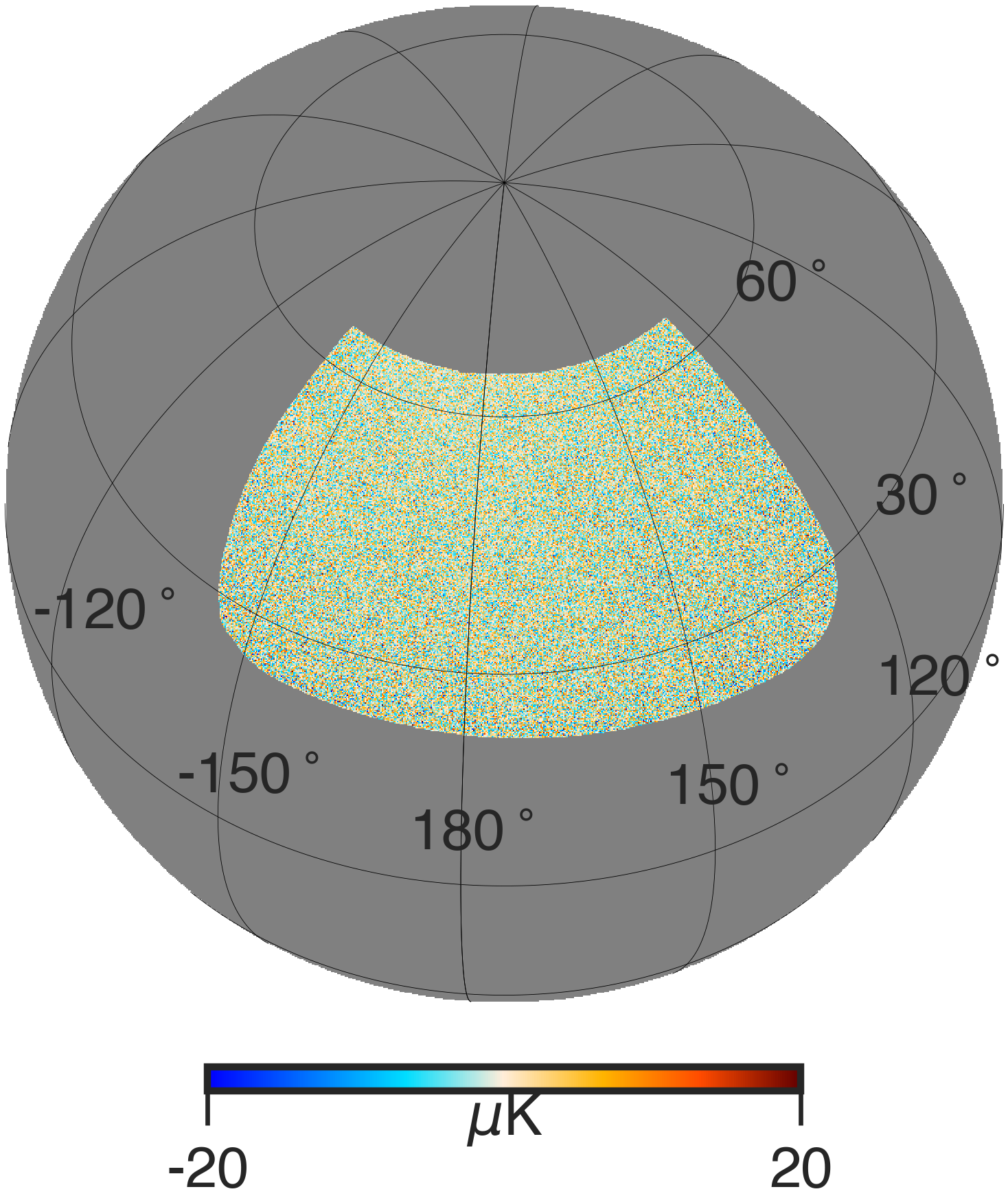}
        \caption{cILC-cleaned $B$ map}
    \end{subfigure}
    \begin{subfigure}[b]{0.3\textwidth}
        \centering
        \includegraphics[width=\textwidth]{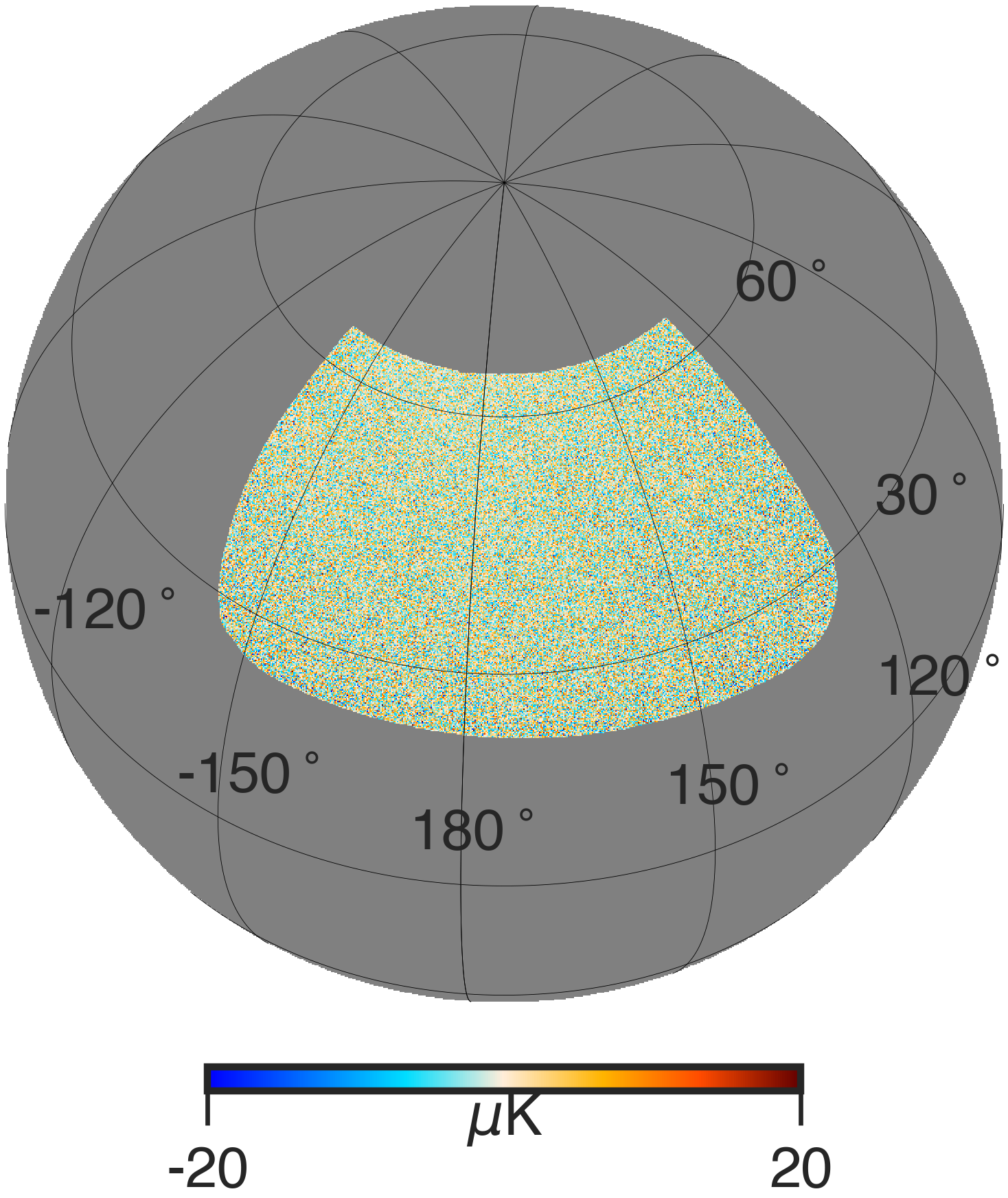}
        \caption{cILC residual $B$ map}
    \end{subfigure}
    \caption{Maps of the input CMB (left column), the foreground-cleaned maps (middle column), and their difference, i.e., the residual maps (right column) for the $T$, $E$, and $B$ modes. The first row shows $T$ maps, the second row shows $E$ maps, and the third row shows $B$ maps.}
    \label{fig:nilc-TEB-maps}
\end{figure*}
\begin{figure*}[htbp]
    \centering
    \begin{subfigure}[b]{0.3\textwidth}
        \centering
        \includegraphics[width=\textwidth]{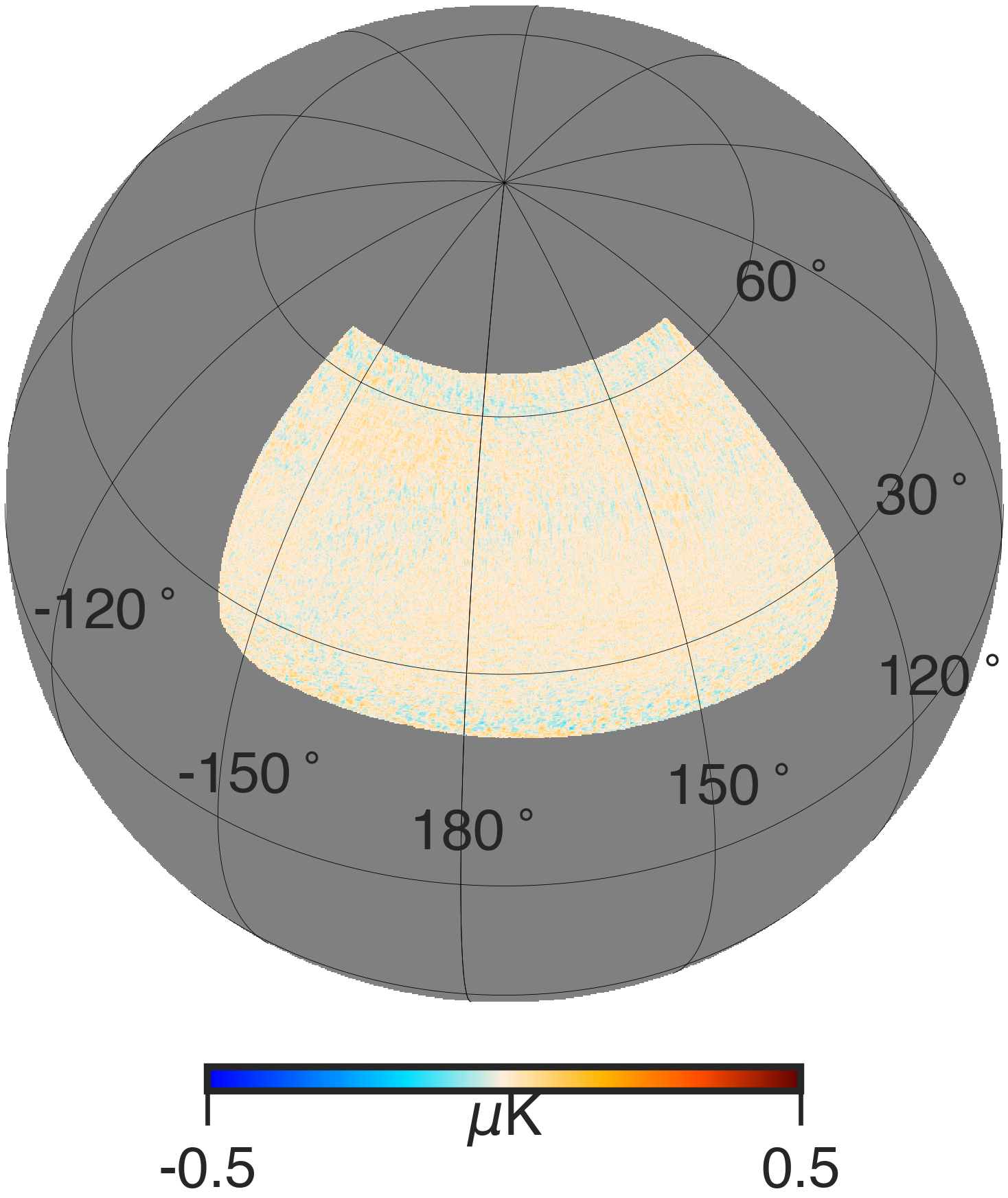}
        \caption{cILC systematic residual}
    \end{subfigure}
    \begin{subfigure}[b]{0.3\textwidth}
        \centering
        \includegraphics[width=\textwidth]{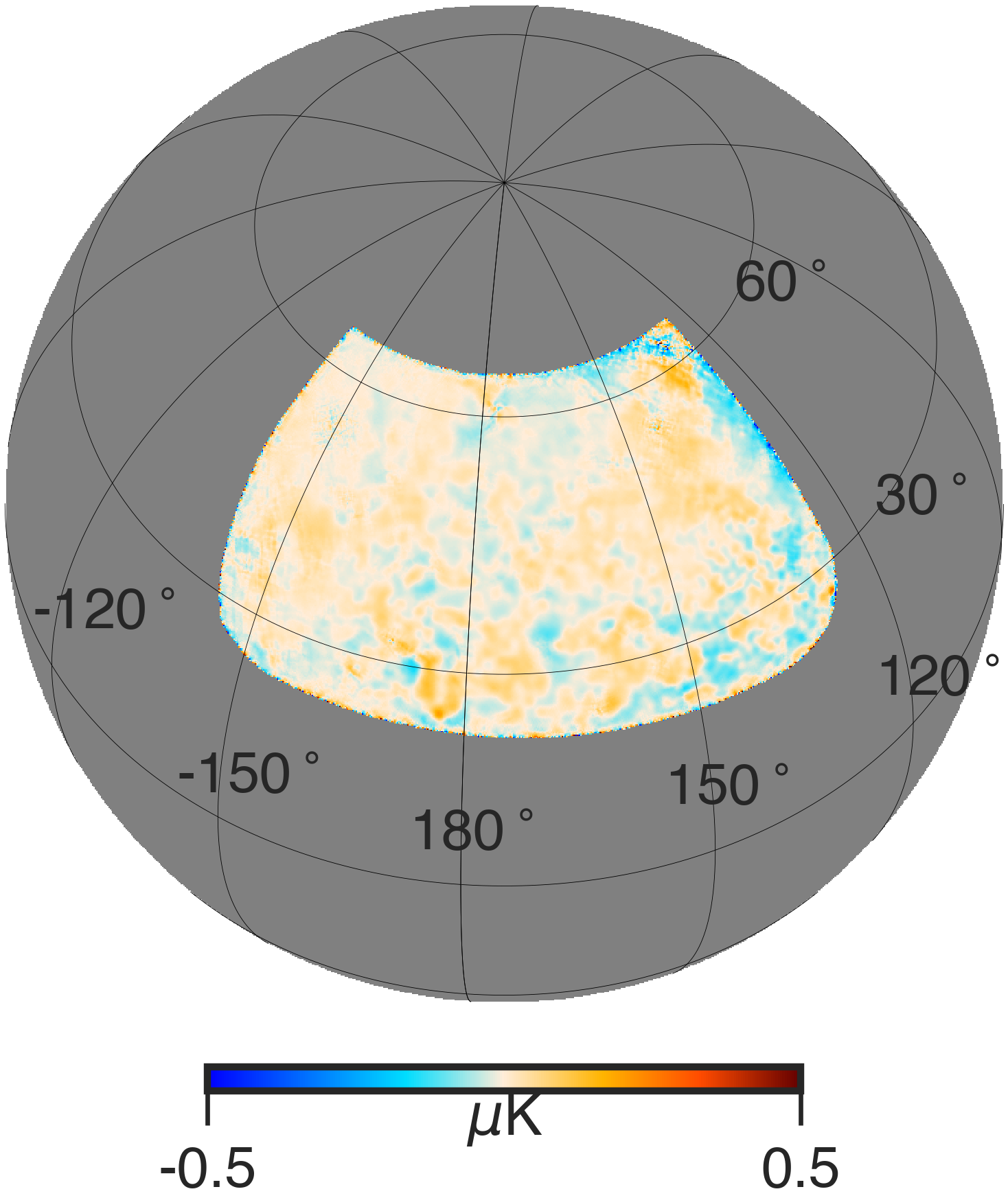}
        \caption{cILC foreground residual}
    \end{subfigure}
    \begin{subfigure}[b]{0.3\textwidth}
        \centering
        \includegraphics[width=\textwidth]{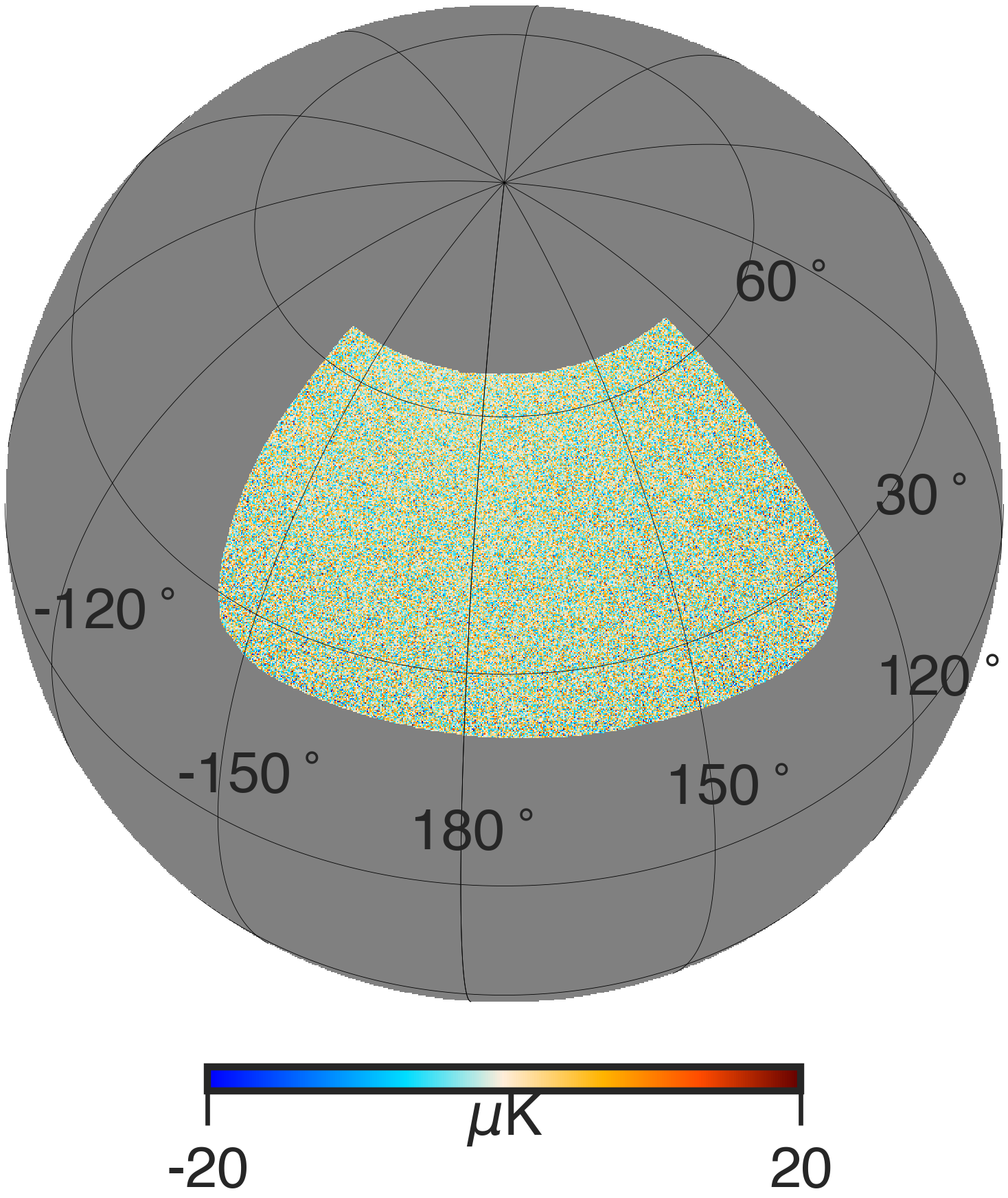}
        \caption{cILC noise residual}
    \end{subfigure}
    \caption{{cILC residual $B$ maps of the systematics (left), foreground (middle), and noise (right) components for one realization.}}
    \label{fig:B-resmaps}
\end{figure*}

\begin{figure*}[htbp]
    \centering
        \includegraphics[width=0.9\textwidth]{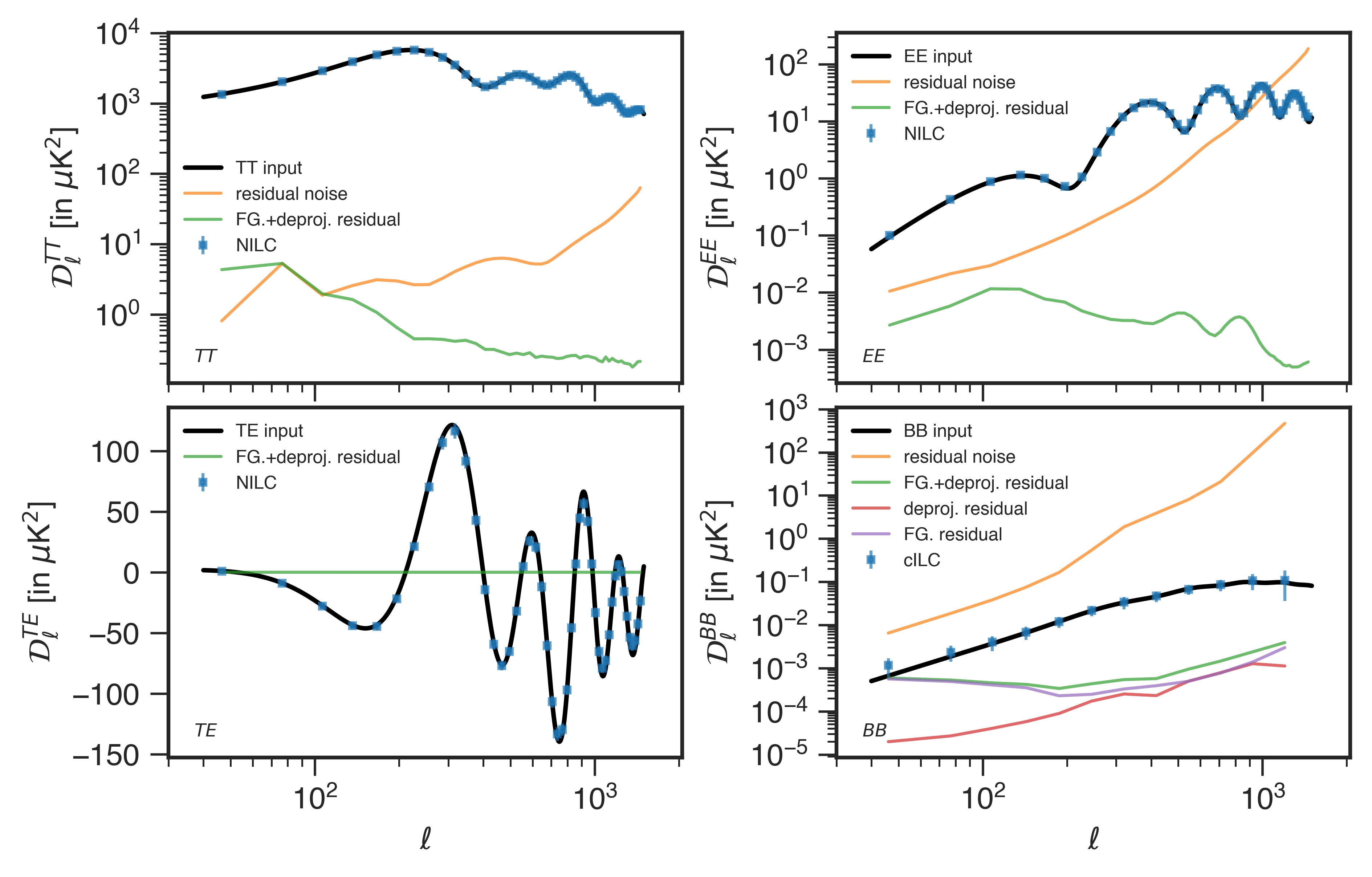}
    \caption{Power spectra including NILC-cleaned $TT$ (left top), NILC-cleaned $EE$ (right top), NILC-cleaned $TE$ (left bottom), and cILC-cleaned $BB$ (right bottom) power spectra {with $\ell_{\max}=1500$}. The input CMB (solid black curves), the reconstructed CMB (blue bars), and the residual noise (orange curves) power spectra are plotted in each panel. The green curves represent the {sum of the systematic and foreground residuals. For the $BB$ spectra, the deprojection and foreground residuals are separately shown as red and purple curves, respectively.} The power spectra are binned over every 30 multipoles except for the $BB$ spectra, where a logarithm binning at $\ell>100$ was adopted for clarity.}
    \label{fig:nilc-power-spec}
\end{figure*}

\section{Results}
\label{sec:res}
In this section, we detail the results of {the fitted beam parameters from deprojection and} the deprojected maps of the S3 dual bands. We then combine the S3 channels with four Planck HFI channels and WMAP K band in the foreground-cleaning pipeline to extract the CMB signal, and we reconstruct the lensing potential power spectrum from the foreground-cleaned maps. {We finally evaluate the impacts of the systematic residual on the CMB power spectra, lensing potential, and the tensor-to-scalar ratio $r$.}

\subsection{{Deprojection results}}
\label{sec:deproj}

{The assumed S3 telescope contains 1728 detector pairs on the focal plane for the 95 and 150 GHz bands, with a field-of-view width $\sim30^\circ$. We considered one observing season (lasting about half a year), which was divided into 3183 scansets (each lasting roughly 30 minutes). A scanset covers tens of constant elevation scans (CES) that the telescope operates from one end to the other at a speed of $\sim4^\circ/s$. Between the scansets, the telescope makes elevation nodding (elnod) scans to calibrate the relative gains of the detectors. We generated time streams from the simulated CMB plus foregrounds following the S3 scan trajectory. We derived the six leakage templates from the Planck simulations as described in Sect.~\ref{sec:bs}. Although the input beam parameters are constant with time, we fit the time-ordered templates to the S3 TOD every 30 scansets for each detector pair to account for the time-varying beam systematics in realistic instruments. The histogram of the fitted beam mismatch parameters of 106 time chunks for one detector pair of a realization at 150 GHz is shown in Fig.~\ref{fig:bsp-150}. The scatter across time slices would be averaged in the map-making process, and the deviation of the mean from the true values (input parameters) due to noise fluctuations in templates leads to the undeprojected residual.}




{We note a systematic bias on the recovered differential plus-ellipticity $\delta p$ at a level of $\Delta=\delta p_{\mathrm recov}-\delta p_{\mathrm input}\approx7.3\times10^{-3}$. Due to the CMB-inherent $TE$ correlation, the deprojection template of $\delta p$ correlates with the polarized signal in the pair-difference time stream, resulting in a bias on the deprojection coefficient even when it is free of beam systematics \citep{bicep2collaborationBICEP2IIIInstrumental2015}. This bias does not impair the filtering of $T\rightarrow P$ leakage, but leads to an additional filtering of cosmological $E$ modes (the effects on the $B$ modes are negligible). We corrected for the bias using the averaged deprojection coefficient estimated from systematics-free simulations, which is $\langle\delta p_{\mathrm recov}\rangle=7.2\times10^{-3}$.}\footnote{{The bias can alternatively be taken into account in the final power spectrum by a suppression factor on the $E$ modes derived from the deprojected systematics-free simulations.}} We also checked the strong correlation between the time-averaged recovered coefficients and the input coefficients for all detector pairs in Appendix~\ref{sec:fit}.

{Finally, we subtracted the fitted templates from the pair difference data for all time trunks of detector pairs. We simplified the map-making process by averaging the time streams of detectors pointed at each map pixel and adding white-noise maps to obtain the observed sky maps.}

\subsection{Frequency maps}
\label{sec:maps-ali}
We performed the deprojection procedure on CMB-plus-foregrounds TOD with beam systematics of the S3 95 and 150 GHz bands {and implemented the map-making to obtain the frequency maps}. The 150 GHz $B$-mode maps before and after the deprojection pipeline are plotted in the first row of Fig.~\ref{fig:B150-maps}, where the left plot shows the input CMB plus foregrounds without beam systematics, the middle plot shows the CMB plus foregrounds with beam systematics, and the right plot shows that after deprojection. The beam systematics with the level considered in this work only substantially affect the $B$ modes, which induce a spurious signal mainly at small scales. The {deprojection residual maps shown in the} second row of Fig.~\ref{fig:B150-maps} illustrate that the deprojection can recover the input {$T$-, $E$-, or $B$-mode} signal. The deprojection has no effect on the $T$ modes, but the noise fluctuation in the deprojection template may lead to a residual $T$-to-$P$ leakage on the $E$ and $B$ modes. However, our results show that the residual leakage almost vanishes in deprojected maps.

We used the PCL-TC estimator {(as detailed in Appendix~\ref{sec:pcl-tc})} to estimate the power spectra of 150 GHz maps, as shown in Fig.~\ref{fig:B150-ps}. {The band powers have a multipole bin size of 30 and $\ell_{\max}=700$.} Only in the $BB$ power spectrum can the beam systematics bias the power spectrum of input CMB plus foregrounds significantly, while the power spectra of deprojected maps for all modes fit the input spectra well. {The residual after deprojection (plotted as the dashed cyan curve) for $BB$ is about 2$\sim$3 orders of magnitude lower than the input CMB plus foreground power spectrum (solid black curves).}



\subsection{Foreground-cleaned maps}
\label{sec:fg-cl-maps}
After propagating the maps with beam systematics through the deprojection pipeline, we used the NILC method to clean the foregrounds of $T$- and $E$-mode maps and used the cILC approach to clean the $B$-mode maps. The cleaned maps are plotted in Fig.~\ref{fig:nilc-TEB-maps}. The left column shows the input CMB maps without adding beam systematics, the middle column shows the NILC-cleaned $T$ and $E$ maps from seven bands, and the right column shows the difference between the cleaned map and the input CMB. The residual maps are dominated by the noise fluctuation for the $B$ modes, while the NILC maps for the $T$ and $E$ modes are dominated by signal.

{For the $B$ modes, we computed the systematic residual map by projecting the cILC weights on the 95 and 150 GHz deprojection residual maps plotted in Fig.~\ref{fig:B150-maps}. In this way, we separated the three components in the $B$-mode cILC residual map from each other, as shown in Fig.~\ref{fig:B-resmaps}. The small-scale artifacts in the systematic contamination map are clearly visible. The residual on the boundaries would be removed through mask apodization in the power spectrum estimation.}




The power spectra of the NILC- and cILC-reconstructed maps are shown in Fig.~\ref{fig:nilc-power-spec}. The orange curves refer to the residual noise power spectrum obtained from projecting the NILC or cILC weights on the input noise maps. The green curves, as mentioned in Sect.~\ref{sec:fg-cl-pip}, represent the sum of the foreground and beam systematic residuals. {For the $B$ modes, the two contaminant components are separately plotted as red and purple curves. The foreground residual dominates the systematic residual at large scales ($\ell\lesssim 300$). The deprojection residual is negligible compared to the CMB $B$-mode power spectrum at all scales.} Our foreground cleaning methods are sufficient to recover the CMB signal, and the residual contaminants remain much lower than the reconstructed CMB power spectrum for the $T$ and $E$ modes and lower than the noise fluctuation for the $B$ modes.



\subsection{Lensing reconstruction}
\label{sec:lens-rec}

In addition to the primordial $B$-mode signal, S3 can also detect the lensing signal with a moderate significance because the noise performance is excellent. In what follows, we evaluate the systematic effects on the lensing reconstruction by using the simulated 4 modules*yr data. 
We fed the 301 simulation sets mentioned above through the lensing pipelines \citep{Carron:2017mqf,Liu2022,Han2023} and tested their performance. 

In the presence of lensing-induced correlations among different multipole moments, the off-diagonal terms of the covariance matrix of the CMB fields become nonzero. This allowed us to calculate the quadratic estimator \citep{Hu:2001kj,Okamoto:2003zw} using pairs of filtered maps in their quadratic form \citep{Carron:2017mqf,Maniyar:2021msb}, as described in Eqs. (3.4) to (3.8) of \citet{Liu:2022beb}. In this approach, one map in the pair is Wiener filtered, while the other is inverse-variance filtered. To mitigate biases introduced by random disconnected noise, masking effects, and foreground residuals, we employed a Monte Carlo simulation method to estimate these contributions and performed a mean-field subtraction by averaging over the quadratic estimators, using 60 simulation sets.

Normalization was conducted in two stages: We first assessed the averaged noise level for the entire mock dataset and computed the normalization factor analytically, assuming an effective isotropic noise level. Then we corrected for normalization bias arising from noise inhomogeneities through numerical simulations. The raw power spectrum of the lensing is simply

\begin{equation}
    \label{eq:phihatpow}
    \hat{C}_L^{\hat\phi\hat\phi}=\frac{1}{(2L+1)f_{\mathrm sky}}\sum_{M=-L}^{L}\hat\phi_{LM}\hat\phi_{LM}^{\ast}\;,
\end{equation}
where $\hat\phi_{LM}$ denotes the harmonic transformation of the reconstructed lensing potential. However, the outcome of the quadratic estimator contains not only the lensing potential signal, but also the Gaussian reconstruction noise sourced by the CMB and instrumental noise ($N_0$ bias) and the nonprimary couplings of the connected four-point function \citep{Kesden:2003cc} ($N_1$ bias). We calculated the realization-dependent $N_{0}$ bias (RDN0) using 240 simulation sets, while the $N_{1}$ bias was calculated analytically. After subtracting these biases, we obtained the final estimated power spectrum,

\begin{equation}
    \label{eq:phipow}
    \hat C^{\phi\phi}_L=\hat C_L^{\hat\phi\hat\phi}-\Delta C_L^{\hat\phi\hat\phi}|_{\mathrm RDN0}-\Delta C_L^{\hat\phi\hat\phi}|_{\mathrm N1}\;.
\end{equation}

\begin{figure}[htbp]
\centering
    \includegraphics[width=10cm]{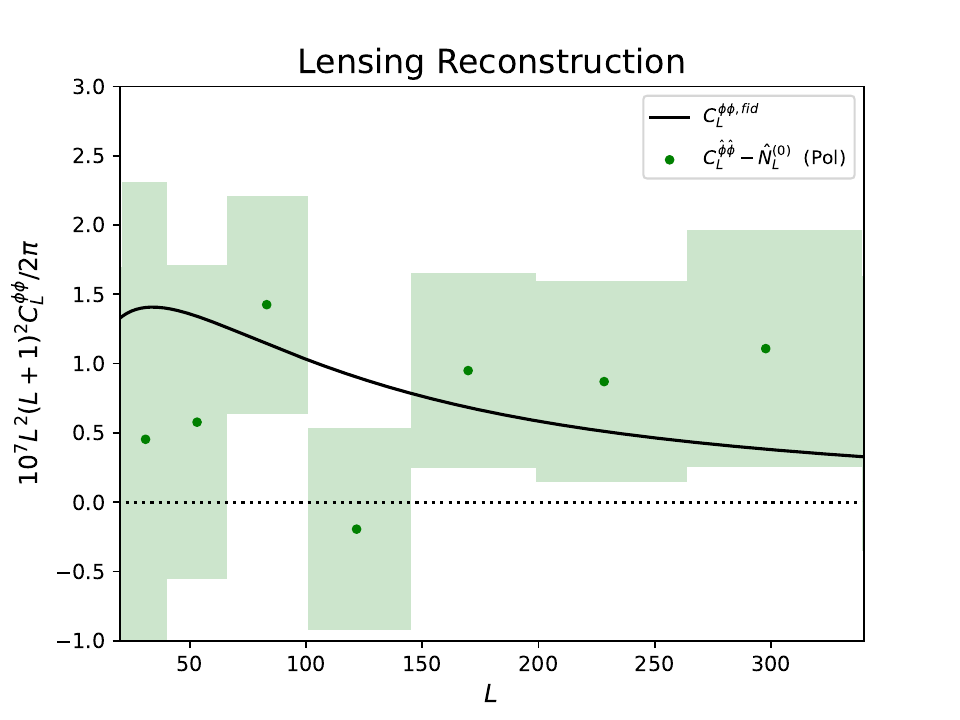}
    \caption{Reconstructed lensing potential power spectrum from the polarization estimator (green points). The colored boxes denote the 1$\sigma$ error regions calculated from 240 simulation sets.}
    \label{fig:recon}
\end{figure}



In Fig.~\ref{fig:recon} we show the reconstructed lensing potential power spectrum from polarization data, which combines the $EE$, $EB$, and $BB$ estimators. The discrepancy in the figure between the reconstructed data point and the theoretical prediction (black curve) originates in the Gaussian uncertainty of the single realization that represents the ``real'' data. Thus, it is expected that some of the data points may deviate by up to $1\sigma$ from the theoretical curve considering that each $\ell$ bin is independent.

To evaluate the outcome of our lensing reconstruction, we estimated the signal-to-noise ratio (S/N) via the Fisher-matrix method,
\begin{equation}\label{eq:SNR}
    {\mathrm S/N}=\sqrt{\sum_{\ell,\ell'}C_{\ell}\mathbb{C}^{-1}_{\ell\ell'}C_{\ell'}}\;,
\end{equation}
and we used the theoretical instead of the reconstruced $C_{\ell}$ in the numerator to ensure a stable prediction. $\mathbb{C}_{\ell\ell'}$ is the covariance matrix obtained from 240 simulation sets, which reads

\begin{equation}\label{eq:Cll}
    \mathbb{C}_{\ell\ell'} = \frac {1}{N-1} \sum\limits_{n=1}^{N=240} \bigg[\Big(\hat C^{\phi\phi}_{\ell} - \overline{C}^{\phi\phi}_{\ell}\Big)\times \Big(\hat C^{\phi\phi}_{\ell'} - \overline{C}^{\phi\phi}_{\ell'}\Big) \bigg] \ ,
\end{equation}
where $\overline{C}^{\phi\phi}_{\ell}$ is the averaged lensing-potential power spectrum based on the simulation sets. The final S/N obtained from the multipole range of $\ell\in(20,340)$ is S/N $\approx4.2$, and this number is consistent with \citet{Han:2023gvr}, where we reported an S/N $\approx4.5$ with the same foreground-removing and lensing- reconstruction pipeline using a database without systematics. We therefore conclude that the systematic residual does not contaminate the lensing reconstruction strongly. 

\subsection{{Tensor-to-scalar ratio bias}}
\label{sec:r-bias}
{In this section, we estimate the bias on the tensor-to-scalar ratio $r$ due to the beam systematic residual by comparing the estimated $r$ bias of the simulations with (after deprojection) and without the beam systematics.}

{We sampled the posterior distribution of the tensor-to-scalar ratio $r$ from the averaged cILC-cleaned $BB$ power spectra by an MCMC analysis. We assumed the Gaussian likelihood of $r$ given the cleaned band powers $\hat C_{\ell_b}$ as}
\begin{equation}
    \begin{aligned}   
        - 2 \ln \mathcal{L}(r) = \sum_{\ell_b \ell_{b'}} &\left[\hat C_{\ell_b} - rC_{\ell_b}^{r=1} - C_{\ell_b}^{\mathrm lens} \right]\left[M^{-1}_{\mathrm fid}\right]_{ \ell_b \ell_{b'}} 
        \\
        &\left[\hat C_{\ell_{b'}} - rC_{\ell_{b'}}^{r=1} - C_{\ell_b}^{\mathrm lens} \right]\,,
    \end{aligned}
    \label{eq:gauss_like}
\end{equation}
{where $\ell_b, \ell_{b'}$ are indices for multipole bins, $rC_{\ell_b}^{r=1} + C_{\ell_b}^{\mathrm lens}$ refers to the theoretical binned $BB$ power spectrum with a multipole range of $\ell\in[40, 200]$ including 5 $\ell_b$, $C_{\ell}^{\mathrm lens}$ is the $B$-mode spectrum due to gravitational lensing, $C_{\ell}^{r=1}$ is the $B$-mode spectrum sourced by tensor perturbations with $r=1$, and $[\boldsymbol M_{\mathrm fid}]_{\ell_b \ell_{b'}} = \langle(\hat C_{{\mathrm fid,}\ell_b} - \langle \hat C_{{\mathrm fid,}\ell_b}\rangle) (\hat C_{{\mathrm fid,}\ell_{b'}} - \langle \hat C_{{\mathrm fid,}\ell_{b'}}\rangle)\rangle$ is the fiducial covariance matrix computed from 300 systematics-free simulations. We derived the posterior distribution function assuming a uniform prior of $r\in[0,1]$. We constructed an MCMC chain consisting of 10,000 samples that satisfied the posterior distribution using the \texttt{emcee} \citep{foreman-mackeyEmceeMCMCHammer2013} Python package. We obtained the 1$\sigma$ confidence interval (C.I.) and the 95\% upper limit of $r$ for the deprojection- and systematics-free cases.}
\begin{figure}[htbp]
\centering 
    \includegraphics[width=\hsize]{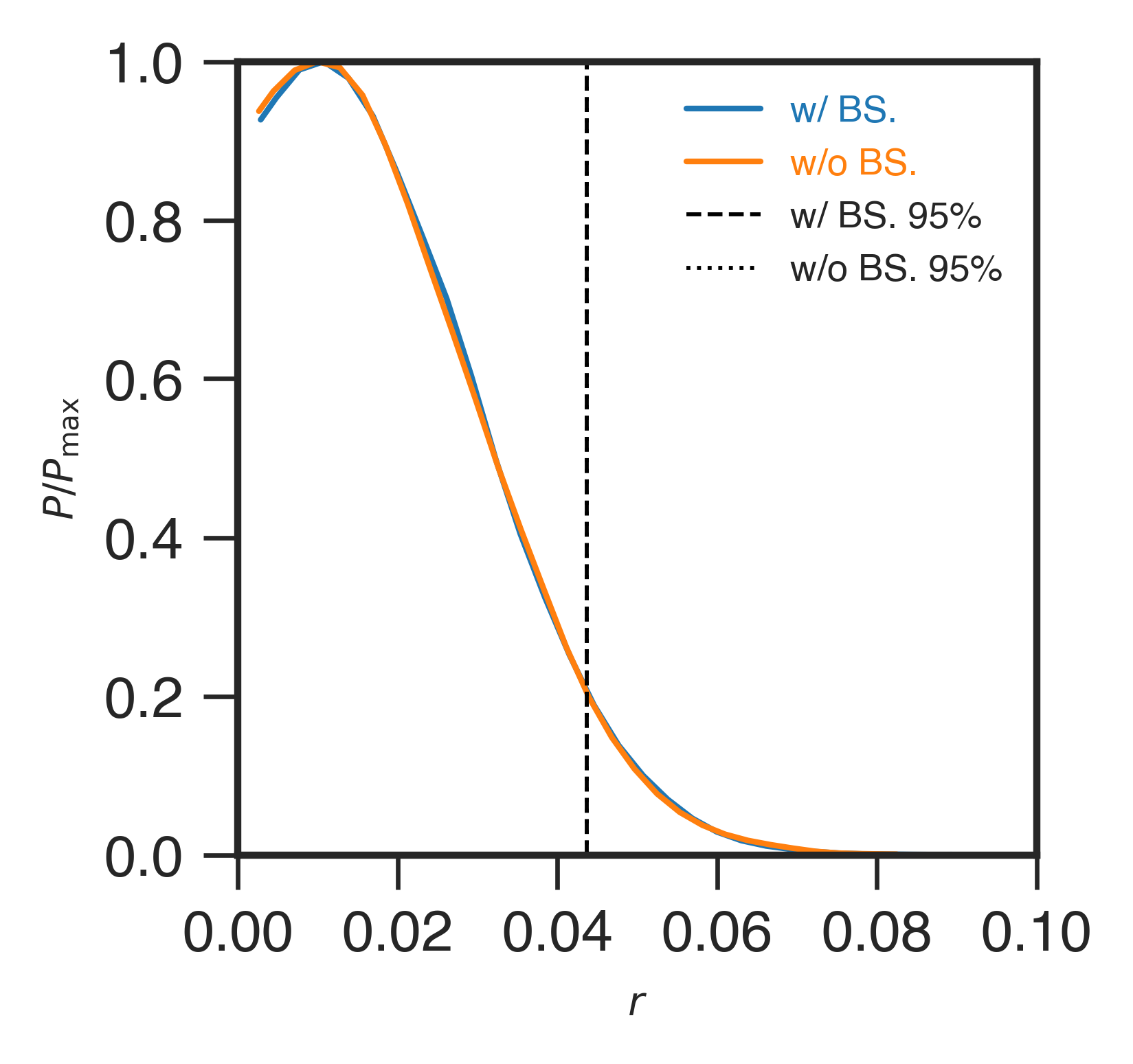}
    \caption{{Distribution of the tensor-to-scalar ratio for the cases with and without beam systematics. The dashed vertical lines represent the 95\% upper bound, which is 0.043 for the two cases.}}
    \label{fig:r-llh}
\end{figure}



{The distribution of $r$ is shown in Fig.~\ref{fig:r-llh}. The resulting 1$\sigma$ C.I. for the systematics-free case is $r=0.019\pm0.013$ and the 95\% C.I. is $r<0.043$, while for the deprojection case, the results are nearly the same. We note that in the selected scales $\ell\in[40, 200]$ where the $B$-mode spectrum is sensitive to tensor perturbations, the deprojection residual is negligible compared to the tensor power spectrum $C_{\ell}^{r=1}$ and foreground contaminants (see Fig.~\ref{fig:nilc-power-spec}). Therefore, the beam systematics imprint negligible effects on the measurements of $r$.}

\section{Conclusions}
\label{sec:concl}
We forecast the systematic effects of beam mismatches on the forthcoming S3 observations. We adopted the deprojection technique in the map-making procedure to mitigate the $T\rightarrow P$ leakage induced by the random beam mismatches added to the S3 mock TOD, and we then assessed the performance of the deprojection on the foreground cleaning, lensing reconstruction, {and $r$-estimation}. First, we compared the CMB-plus-foregrounds after deprojection and the input CMB-plus-foregrounds without instrumental effects in terms of the $TEB$ maps and the angular power spectra. The CMB and foregrounds are mixed up in deprojection and thus cannot be separated from each other. We found that the residual leakage can be ignored compared to the input CMB-plus-foregrounds.

Then, we imposed the foreground-cleaning methods, NILC for $T$ {and} $E$ modes and cILC for $B$ modes, on the 300 simulations of seven frequency bands including the S3 bands, the Planck HFI bands, and the WMAP K band, where the S3 mock data are deprojected. The power spectrum of the biases consisting of the residual $T\rightarrow P$ leakage and foregrounds was shown to be negligible compared to the noise uncertainties for the $TT$, $EE$, $TE$, or $BB$ power. We evaluated the effect of beam systematics on the lensing reconstruction. Our lensing-potential power spectrum reconstructed from the polarization estimator has a similar signal-to-noise ratio {as that obtained in previous analyses} without systematics, which demonstrates that the lensing reconstruction is almost unaffected by residual systematics. {Finally, we set constraints on the tensor-to-scalar ratio $r$ using the cleaned $BB$ spectrum, where we obtained similar results with and without beam systematics because the deprojection technique mitigates the systematics to a negligible level.} Our results justify the use of a deprojection to filter the $T\rightarrow P$ leakage due to beam mismatches in S3 experiments, and the residual leakage can be ignored in the following data analysis pipeline. {The effects of far sidelobes are not involved in this work since they are related to the specific shielding system of the telescope \citep{bicep2BICEP2KeckArray2015} and the calibration or modeling of the realistic beam maps \citep{Gallardo_2018}.  The undeprojected residual due to the differential beams of the sidelobes will be calibrated in future works.} 

\begin{acknowledgements}
We thank Zirui Zhang, Siyu Li and Jing Jin for helpful discussions. This work is supported by the National Key R\&D Program of China Grant No. 2021YFC2203102, 2020YFC2201603, NSFC No. 12325301 and 12273035. Jiakang Han thanks Stefano Camera for various support for this project. 
Some of the results in this paper have been derived using the pymaster, emcee, healpy, and HEALPix packages.
\end{acknowledgements}

\bibliography{Bib}
\bibliographystyle{aa}

\begin{appendix}
\onecolumn

\section{Recovered beam parameters versus the input ones}
\label{sec:fit}

\vspace{-1.2em}
\begin{figure*}[htbp]
    \centering
        \includegraphics[width=0.78\textwidth]{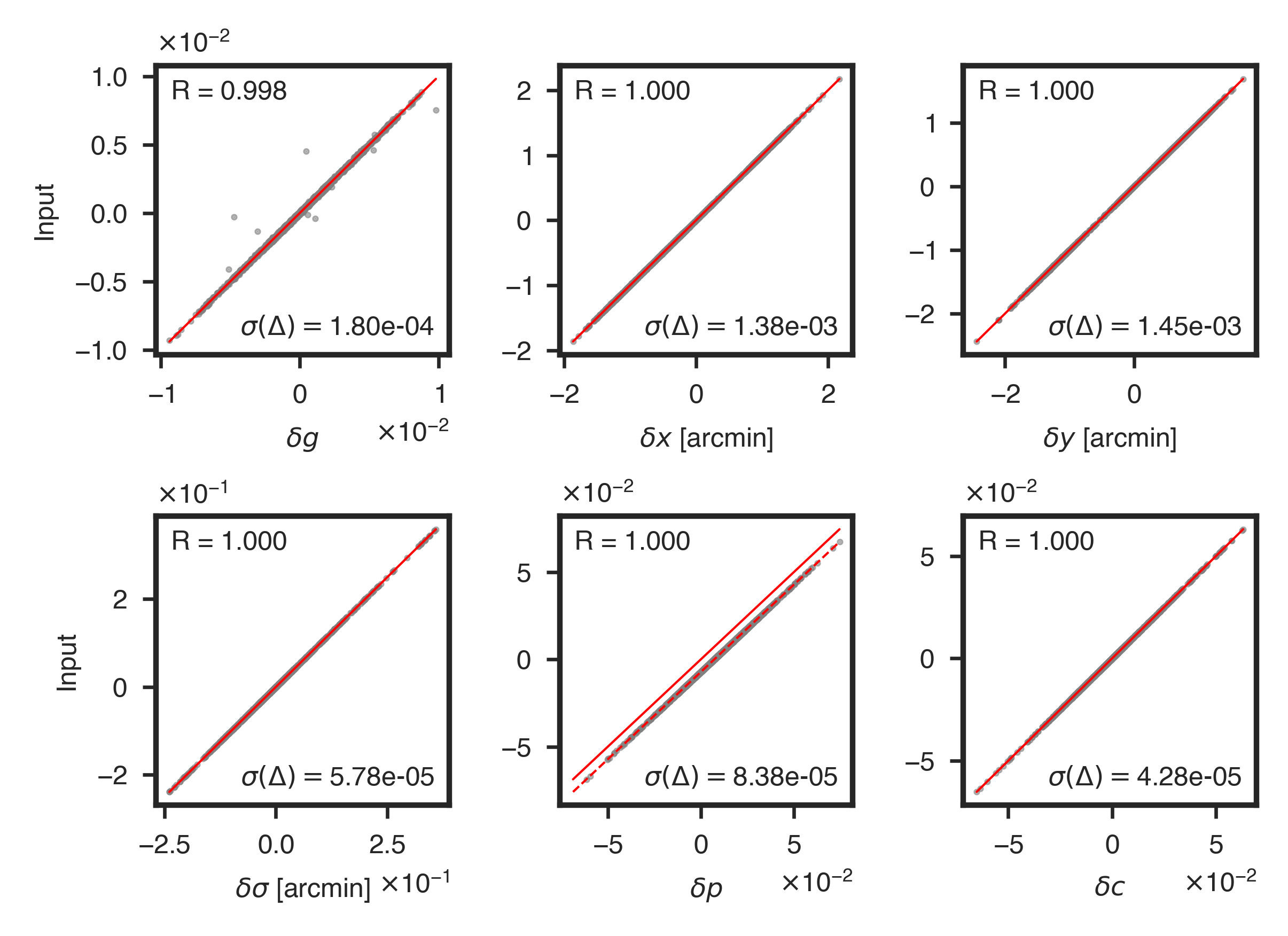}
    \caption{{Scatter plot of the recovered beam mismatch parameters averaged over time trunks ($x$-axis) vs. the input values ($y$-axis) for all 1728 detector pairs of a simulation at 150 GHz. The red solid curves have a slope of 1 and a $y$-intercept of 0. The red dashed curve has been offset vertically by the mean bias on $\delta p$ estimated from systematics-free simulations. The correlation coefficient $R$ and the standard deviation of the bias on the deprojection coefficient $\sigma(\Delta)$ are exhibited.}}
    \label{fig:dpr-150}
\end{figure*}

{Figure~\ref{fig:dpr-150} shows the comparison between the fitted beam mismatch parameters in deprojection and the input values for 1728 detector pairs of a realization at 150 GHz. The red solid curves refer to $y=x$, and the red dashed curve was vertically offset by the estimated mean bias on $\delta p$ ($7.2\times10^{-3}$). The Pearson correlation coefficient $R$ approaching 1 for all parameters indicates a strong correlation between the deprojection coefficients and the input parameters, demonstrating the precision of deprojection. The standard deviation of the bias ($\Delta$) on each coefficient is significantly lower than the median measurement uncertainty for individual detectors by BICEP2/Keck Array \citep[see][Table~3]{bicep2BICEP2KeckArray2015}, but is too optimistic since the noise was not added to the data.}

\FloatBarrier

\section{{PCL-TC estimator}}
\label{sec:pcl-tc}
{The $E\rightarrow B$ mixing is introduced when the partial-sky $Q$ and $U$ maps are converted to the $E$ and $B$ modes in harmonic space \citep{zhaoSeparatingTypesPolarization2010,ghoshEndingPartialSky2021}. We used the template cleaning method \citep{liuBlindCorrectionEBleakage2019,liuMethodsPixelDomain2019} to get rid of the resulting $E\rightarrow B$ leakage. To estimate the full-sky $B$-mode power spectrum from foreground-cleaned maps, we adopted the pseudo-$C_\ell$ method to account for mode-coupling induced by masking and multipole binning using the \texttt{NaMASTER} \citep{alonsoUnifiedPseudoC_2019}\footnote{https://github.com/LSSTDESC/NaMaster} package. The whole process is called the ``PCL-TC'' estimation.}

{The template cleaning method is a pixel-based technique to remove $E\rightarrow B$ leakage due to partial-sky effects. The steps are as follows:}
\begin{enumerate}
    \item {First convert the masked $(Q,U)$ maps into $(a_{\ell m}^E, a_{\ell m}^B)$ using the \texttt{map2alm} function of \texttt{healpy} \citep{Zonca2019}. Transform $a_{\ell m}^B$ into the $B$ map being contaminated by the $E\rightarrow B$ leakage, $B_c$.}
    \item {Convert the $E$-only modes $(a_{\ell m}^E, 0)$ back into maps $(Q_E,U_E)$ using the \texttt{alm2map} function of \texttt{healpy}. Multiply them by the mask and convert them into $(a_{\ell m}^{E\prime}, a_{\ell m}^{B\prime})$ using \texttt{map2alm}. The $B$ map derived from $a_{\ell m}^{B\prime}$ is taken as our $E\rightarrow B$ leakage template, $B_t$.}
    \item {Linearly fit the masked leakage template $B_t$ to the masked contaminated map $B_c$ in the pixel space. Finally subtract the fitted leakage to obtain the pure $B$-mode map: $B_{\mathrm pure}=B_c-\alpha B_t$ where $\alpha$ is the linear coefficient.}
\end{enumerate}

{Deprojection, as any other filtering operation on $Q$ and $U$ time streams, can produce a small amount of $E\rightarrow B$ mixing. The $E\rightarrow B$ leakage due to deprojection can also be corrected for by taking advantage of the deprojected $E$-mode only data to get a $B$-mode leakage template \citep[see][Sect.~4.2]{ghoshPerformanceForecastsPrimordial2022}.}

{After the foreground cleaning, we corrected for the common beam (11 arcmin) and pixel window functions in the $a_{\ell m}$'s. Finally we computed the full-sky bandpowers with a bin size of 30 using the \texttt{decouple\_cell} module of \texttt{pymaster}.}

\twocolumn

\end{appendix}

\end{document}